\documentclass[12pt]{article}

\usepackage{graphicx}
\usepackage{latexsym,natbib}
\usepackage{amssymb}

\usepackage[table]{xcolor}

\usepackage[hidelinks]{hyperref}

\newcommand{\bea}{\begin{eqnarray*}}
\newcommand{\eea}{\end{eqnarray*}}
\newcommand{\beq}{\begin{equation}}
\newcommand{\eeq}{\end{equation}}

\newcommand{\cb}{\cellcolor{blue!25}}
\newcommand{\ccr}{\cellcolor{red!25}}
\newcommand{\cy}{\cellcolor{yellow!50}}

\def\bc{\begin{center}}
\def\ec{\end{center}}

\def \R {\mathchoice {\hbox{I\kern -0.1667em R}}{\hbox{I\kern -0.1667em R}}
                      {\small{I\kern -0.1667em R}}{\small{I\kern -0.1667em R}}}

\begin{document}

\begin{center}

{\Large Analyzing genome-wide association studies with an FDR controlling modification of the Bayesian information criterion}\\[6mm]

  $^{1}$Erich Dolejsi, $^{2}$Bernhard Bodenstorfer, $^{1}$Florian Frommlet
  \\

 \bigskip

{\it  $^1$ Section of Medical Statistics, CEMSIIS, Medical University Vienna, Austria\\
$^2$ cogiti e.U., Karlstift, Austria
}
\end{center}
\bigskip

\begin{abstract}

 The prevailing method of analyzing GWAS data is still to test each marker individually, although from a statistical point of view it is quite obvious that in case of complex traits such single marker tests are not ideal. Recently several model selection approaches for GWAS have been suggested, most of them based on LASSO-type procedures. Here we will discuss an alternative model selection approach which is based on a  modification of the Bayesian Information Criterion (mBIC2) which was previously shown to have certain asymptotic optimality properties in terms of minimizing the misclassification error. Heuristic search strategies are introduced which attempt to find the model which minimizes mBIC2, and which are efficient enough to allow the analysis of GWAS data.

 Our approach is implemented in a software package called MOSGWA.  Its performance in case control GWAS is compared with the two algorithms HLASSO and GWASelect, as well as with single marker tests, where we performed a simulation study based on real SNP data from the POPRES sample. Our results show that MOSGWA performs slightly better than HLASSO, whereas according to our simulations GWASelect does not control the type I error when used to automatically determine the number of important SNPs. We also reanalyze the GWAS data from the
Wellcome Trust Case-Control Consortium (WTCCC) and compare the findings of the different procedures.

The program is available at  \href{http://mosgwa.sourceforge.net/}{http://mosgwa.sourceforge.net/}. An application note describing the software in more detail is in preparation. 

The data set used for simulations in this manuscript was obtained from dbGaP through dbGaP accession number phs000145.v1.p1 at 
{\scriptsize www.ncbi.nlm.nih.gov/projects/gap/cgibin/study.cgi?study\_id=phs000145.v1.p1.}

 The WTCCC data is available at http://www.wtccc.org.uk/.

\end{abstract}

\section{Introduction}

Recently there has been growing interest in model selection
approaches to GWAS analysis. Although it is still common practice in
published GWAS to perform statistical analysis for each SNP
individually, there is increasing awareness that this kind of single
marker analysis has certain deficiencies in case of complex traits.
Several authors have commented that marginal tests will suffer from
lack of power to detect SNPs because the effect of other causal SNPs
remains unaccounted for (see for example \cite{Hog8, FRTB}). It has
been argued that this shortcoming of single marker tests might play
a significant role in the widely discussed phenomenon of ``missing
heritability'' in GWAS \citep{Yang10}.

A slightly more sophisticated and less known problem has been
pointed out by \cite{FRTB}, namely that single marker tests have
serious difficulties to rank important SNPs correctly.  This is
obvious for SNPs which are not directly associated with a trait, but
which have an important effect conditional on the presence of other
SNPs. However, even in case of SNPs with marginal effects it turns
out that due to small sample correlations some important SNPs might
have rather small probability to be detected, whereas other SNPs
which are not associated at all with the trait might be selected
with large probability. This result puts in question the common
practice to report those SNPs in GWAS which have lowest ranking
marginal p-values.

Given these deficiencies of single marker tests one can expect that
the use of multi marker models to analyze GWAS will become more and
more important. Multiple linear regression models for quantitative
traits and logistic regression models for case control studies have
a long history in genetic association studies. To facilitate their
use for GWAS there is a strong demand of two things: A thorough
theoretical understanding of different model selection strategies in
high dimensions to find the regression model which includes
important SNPs, as well as the availability of software packages
which make modern statistical methodology applicable to GWAS
analysis.

Concerning the theory of high dimensional data analysis the last two decades have seen a large number of innovations. One milestone was the development of LASSO by \cite{Tib96}, which paved the way for a large number of other new approaches to model selection. A comprehensive presentation of the theoretical foundations of LASSO and its many extensions like adaptive LASSO, group lasso or the elastic net can be found in \cite{BV11}. In the context of GWAS several algorithms have been implemented based on  LASSO or one of its extensions \citep{Koo, He11, Wu} .

From a Bayesian perspective the LASSO is equivalent to model selection with a double exponential (DE) distribution as shrinkage prior. Among the first software packages which allowed to perform multi marker analysis of GWAS was HLASSO \citep{Hog8}, which uses not only DE priors, but alternatively considers normal exponential Gaussian (NEG) priors. The NEG distribution is more pointed than DE at 0, resulting in the selection of potentially smaller models.  More recently another version of the Bayesian LASSO for GWAS analysis was introduced by \cite{Li11}.

The LASSO itself was originally developed for model selection
problems of moderate size, whereas in GWAS one usually is confronted
with up to a million SNPs or more. For such ultra-high dimensional
problems \cite{Fan8} suggested sure independence screening (SIS) as
a convenient way of dimension reduction. In case of regression
models SIS is nothing else but preselecting a certain number of
markers based on marginal tests.  After SIS more refined methods
like LASSO or SCAD can be applied to obtain a model. Using tests
conditional on that selected model over all remaining markers one
can apply another SIS step. Iterating SIS and refined model
selection gives the procedure called ISIS.

A startlingly simple but computationally intensive method to improve
the performance of model selection procedures in high dimensions is
stability selection introduced by \cite{MB10}. In stability
selection random subsamples of the data are drawn repeatedly and a
given model selection procedure like LASSO is performed on each of
these subsamples. The final model is then obtained by considering
those regressors which have appeared consistently over the repeated
samples. 

One of the more prominent model selection packages for GWAS
is GWASelect by \cite{He11}, which combines ISIS with stability
selection based on 50 random subsamples, where the refined model
selection procedure in ISIS is LASSO.  
GWASelect itself uses a prespecified size of the model, but there exists a 'dynamic'
version d-GWASelect which uses cross-validation to fit the LASSO parameter and which
determines the number of selected SNPs using stability selection.  
Thus in this article  GWASelect actually refers to d-GWASelect, the
algorithm which  allows to determine the number of interesting SNPs.

An alternative approach to model selection in high dimensions which
is currently gaining popularity is based on information criteria.
Among the large number of SNPs genotyped in GWAS one expects only a
moderate number of SNPs to have a strong effect. In such a sparse
setting classical information criteria like Akaike's AIC or
Schwarz's Bayesian information criterion (BIC) have been shown to
select too large models \citep{BS}. Consequently \cite{BGD}
introduced a modification of BIC called mBIC which is designed to
control the family wise error rate (FWER) of selected markers in
sparse regression \citep{BZG}. A rather similar criterion called
EBIC was presented by \cite{Chen1}, who showed consistency results
for EBIC under sparsity even in case when the number of markers is
growing faster than the number of observations.

Recently \cite{BCFG} coined the notion of asymptotic Bayes
optimality under sparsity (ABOS) which is a stronger property than
model selection consistency. \cite{FBMC} introduced several further
modifications of BIC which have the property of controlling the
false discovery rate (FDR, see \cite{BH}), among them mBIC2. It was
shown that the FWER controlling mBIC is ABOS only in case of extreme
sparsity, whereas the FDR controlling criteria are ABOS under a much
wider range of sparsity levels. Further background information on
mBIC2 is given by \cite{FRTB}. In the context of quantitative traits
extensive GWAS simulations based on real SNP data show that mBIC2 is
considerably more powerful to detect causal SNPs than mBIC, while
controlling the FDR at a fixed level.

In this article we will focus on case control studies, where the model selection task is performed using logistic regression models. Computing maximum likelihood estimates for each model is much more time consuming than in case of quantitative traits, and we therefore developed rather involved search strategies trying to find models which minimize mBIC2.  The resulting algorithm is implemented in the software package MOSGWA, and we compare its
performance with single marker tests and with other variable selection methods, in particular
with HLASSO \citep{Hog8} and GWASelect \citep{He11}. The main reason
for this choice is that \cite{He11} already performed a comparison
with several other methods for GWAS analysis, and HLASSO and
GWASelect gave the most convincing results.

The rest of this paper is organized as follows. In Section
\ref{Sec:Methods} we give a detailed description of the algorithmic aspects of 
MOSGWA. Section \ref{Sec:Simulations} describes our
simulation scenarios for case control studies and compares the
performance of different procedures. Then we reanalyze the GWAS data from the Wellcome Trust Case-Control Consortium (WTCCC), where Section \ref{Sec:RealData} gives a summary and Section \ref{Sec:WTCCC_Detail} a detailed exposition of the results of this analysis.  Finally in
Section \ref{Sec:Discussion} we discuss our findings and point out
how to extend MOSGWA in the future.

\section{Methods}
\label{Sec:Methods}

\subsection{Selection criterion}
Before describing the algorithmic details of
MOSGWA we will introduce the selection criterion mBIC2. Consider a GWAS  based on $p$ SNPs and $n$ individuals.
For a given model including $k_M$ SNPs our model selection criterion is of the form
\beq
\label{mBIC2} mBIC2 =
-2 \log L_M^* + k_M \log (n p^2 /4) - 2 \log (k_M!) \; .
\eeq 
Each set of SNPs forms a potential model $M$, and MOSGWA tries to find
that model which minimizes mBIC2. Here $L_M^*$ is the Firth corrected maximum
likelihood, which will be discussed in more detail below.

The penalty of the mBIC2 criterion was introduced in \cite{FBMC}, where its derivation
was based on ideas of \cite{ABDJ}. In particular mBIC2 is closely
related to the Benjamini-Hochberg multiple testing procedure
\citep{BH}, and it controls the false discovery rate of detected SNPs.
In the context of linear regression \cite{FBMC} have  shown certain asymptotic optimality properties of mBIC2.
 Roughly speaking it minimizes the misclassification error when both $p$ and $n$ are large,
 while the number of regressors of the correct model is relatively small.

An extensive motivation of mBIC2 can be found in \cite{FRTB}, where the
criterion is applied for linear regression models to analyze GWAS 
with quantitative traits. In contrast we will focus here on
case-control studies, and just like \cite{He11} and \cite{Hog8} we
make use of logistic regression to model the disease risk of SNPs. 
To this end let $Y_i, i = 1,\dots,n$ denote the disease status of an individual ($Y_i = 1$ for a case, 
$Y_i = 0$ for a control), and let $x_{ij}$ denote the genotype of SNP $j \in\{1,\dots,p\}$ for 
individual $i$. If a model $M$ includes the SNPs $j_1, \dots, j_k$  then the corresponding logistic regression 
model can be written as
\beq \label{logreg}
\pi_i := P(Y_i = 1|M,\theta) = \frac{\exp(\beta_0 + \sum\limits_{r=1}^k \beta_r x_{ij_r})}{1 + \exp(\beta_0 + \sum\limits_{r=1}^k \beta_r x_{ij_r})} \;,
\eeq
with the parameter vector $\theta = (\beta_0,\dots, \beta_k)^T$.  These $k+1$ parameters can be routinely estimated by  
maximizing the corresponding likelihood  $L_M(\theta)$, although occasionally the well known problem of separation may occur, where some
parameter estimates tend towards infinity \citep{AA}. In classical statistical applications where $p \ll n$ 
separation typically arises only in case of small sample sizes. For
GWAS the sample size is usually very large, but the number of
potential regressors is even several orders larger, which results in
many combinations of SNPs for which separation occurs.
\cite{Hei02} suggested to overcome the problem of separation using a bias
corrected version of logistic regression which was originally
proposed by  \cite{Firth93}. The likelihood of the logistic
regression model is multiplied with the corresponding Jeffreys
prior, which is just the square root of the determinant of the Fisher information matrix $I(\theta)$. 
Thus the Firth corrected maximum likelihood from equation (\ref{mBIC2}) is given by
\beq  \label{Firth}
L_M^*  = \max\limits_\theta  L_M(\theta) \sqrt{|I(\theta)|}  \; ,
\eeq
and explicit formulas are given for example in \cite{Hei02}. The Firth-corrected log-likelihood $\log L_M^*$ includes the penalty term  $\log \sqrt{|I(\theta)|}$, which guarantees that parameter estimates cannot get excessively large.  
Note that LASSO based procedures like GWASelect do not run into difficulties with separation because the $L^1$-penalty yields automatically a shrinkage of parameters.

\subsection{Search strategy} \label{Subsec:Search}

Having defined the model selection criterion (\ref{mBIC2}) the main task is to find the model which minimizes mBIC2. The resulting problem is an extremely challenging mixed integer program, for which one can only attempt to develop heuristic methods which yield an approximate solution. The search algorithm of the software package MOSGWA repeatedly makes use of a strategy called fast stepwise search (FSS). 

The aim of FSS is, starting from some initial model, to perform a search heuristic which finds a model with a smaller value of a given selection criterion.  The final call of FSS is performed with the target criterion mBIC2, but within the search it is valuable to work with less stringent criteria to avoid getting stuck in local minima corresponding to models which are missing some of the causal SNPs. Specifically we consider the milder criterion
$$
mBIC_{60} := -2 \log L_M^* + k_M \log (n p^2 /60) \; . 
$$
FSS depends on a pre-specified order of all markers not included in the initial model. This order is either based on some marginal test statistics, or on some conditional score tests as described below.

We will formally write the fast stepwise search as a function 
$$
M = \mbox{FSS}(M_{init}, test, criterion) \; , 
$$ 
to emphasize that it depends on the initial model, on the specific order of markers according to \emph{test}, and on the respective \emph{criterion}. Depending on the order of markers two groups are considered:  Group $G_1$
consists of the best $m_1$ SNPs, and group $G_2$ of the best $m_2$ SNPs. Thus $G_1 \subset G_2$, where $G_1$ is the set of SNPs along which specifically directed forward steps are performed (see below), whereas SNPs within $G_2$ might enter the model via so called exchange steps (see below). The exact choice of the parameters $m_1$ and $m_2$ turns out to be not too important. The default values of MOSGWA which are also used for simulations are $m_1 = 350$ and $m_2 = 5000$ (as long as $p\geq 5000$).  

FSS is based on three algorithmic steps which we call directed forward, exchange, and backward step. FSS starts with considering the initial model $M_{init}$ as the current model. The directed forward step repeatedly tests if enhancing the current model with a SNP decreases the \emph{criterion}, where SNPs within $G_1$ are considered in the order obtained from the \emph{test} (therefore \emph{directed} forward search). The first SNP which improves the current model is added, and an exchange step follows.

In the exchange step all SNPs in the current model are tested whether exchanging them with suitable other SNPs decreases the \emph{criterion}. Suitable candidates for exchanging SNP $S_i$ are all other SNPs  within $G_2$ whose physical distance to $S_i$ on the chromosome is less then $d = 50$. The idea behind this strategy is that in the directed search step it might happen that not the optimal SNP was chosen, but a correlated SNP might further improve the model. Also after several SNPs have been added to the model it can happen that exchanging a particular SNP of the model is beneficial. Limiting the exchange to SNPs close to $S_i$ which themselves have reasonably large test statistic makes this strategy computationally feasible.

The third step of FSS is an extended backward elimination step. A standard greedy elimination step is performed, and if this does not improve the model greedy elimination is repeated up to three times to look for better models. The resulting best model is then starting point for another directed forward step. Directed forward, exchange, and backward steps are then performed repeatedly till no further improvement of the is \emph{criterion} achieved.

Now starting with the null model $M_0$ the complete search strategy of MOSGWA can be specified as follows:
\begin{enumerate}
\item $M^*$ = FSS($M_0$, Cochran Armitage, mBIC\_{60})
\item $M^{**}$ = FSS($M^*$, Score Test, mBIC\_{60})
\item $M^{\mbox{final}}$ = FSS($M^{**}$, Score Test, mBIC2)
\end{enumerate}

The general strategy of this algorithm can be motivated as follows. In the first round markers are preselected based on their marginal (Cochran Armitage) test statistic. In the second round score tests conditional on the model $M^*$ of the first round are performed over all remaining SNPs as described in \cite{He11}. Score tests have the benefit of being computationally much less expensive than likelihood ratio test, and therefore provide a fast way to preselect markers which might be of importance additional to markers within $M^*$. 

The first two rounds are performed with the milder criterion mBIC\_{60}, which is expected to yield models which are too large. In fact when models are getting too large then the value of $d$ from the exchange step is reduced to guarantee reasonable runtime. As mentioned previously the benefit of first working with mBIC\_{60} is that one reduces the chances of missing out on important SNPs due to local minima. Only in the final round FSS is performed with the target criterion mBIC2, and should then yield a model for which the type I error rate is controlled in terms of FDR.

\section{Simulation Studies}
\label{Sec:Simulations}

\subsection{Global null hypothesis} \label{Sec:null}
\begin{figure}[!b]
\label{fig:total_null}
\caption{Illustration of the simulation results under the total null corresponding to Table  \ref{Tab:Null}. The average number of false positives for MOSGWA is compared with HLASSO (left panel) and with GWASelect (right panel), for which false positives were clustered. Simulations were performed for four different numbers of chromosomes, resulting in different numbers of SNPs plotted on the x-axis. 
}
\vspace{-2cm}
\centering
  \begin{minipage}{6.6cm}
    \begin{center}
      \includegraphics[scale=0.33]{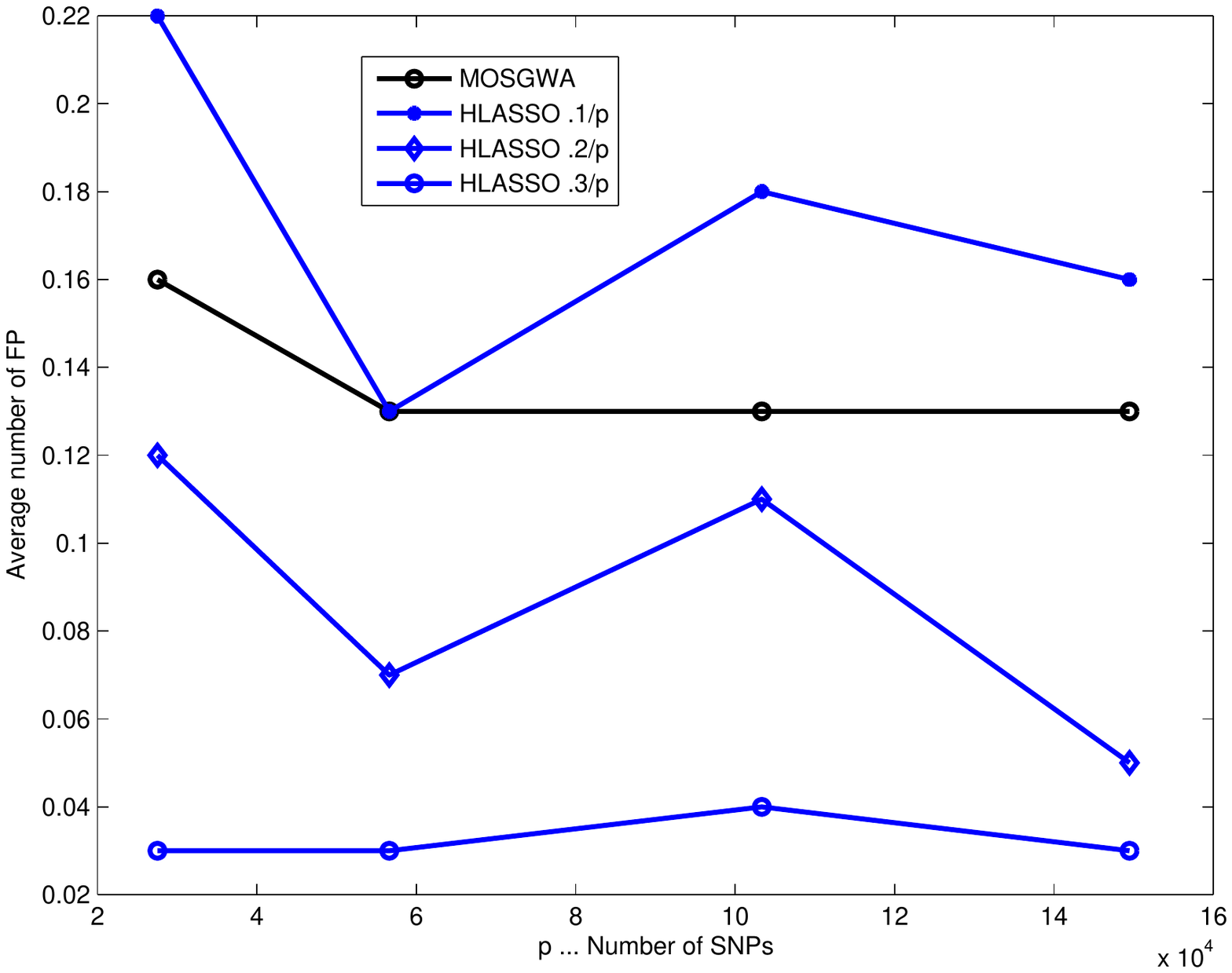}
    \end{center}
  \end{minipage}
    \begin{minipage}{6.6cm}
    \begin{center}
      \includegraphics[scale=0.33]{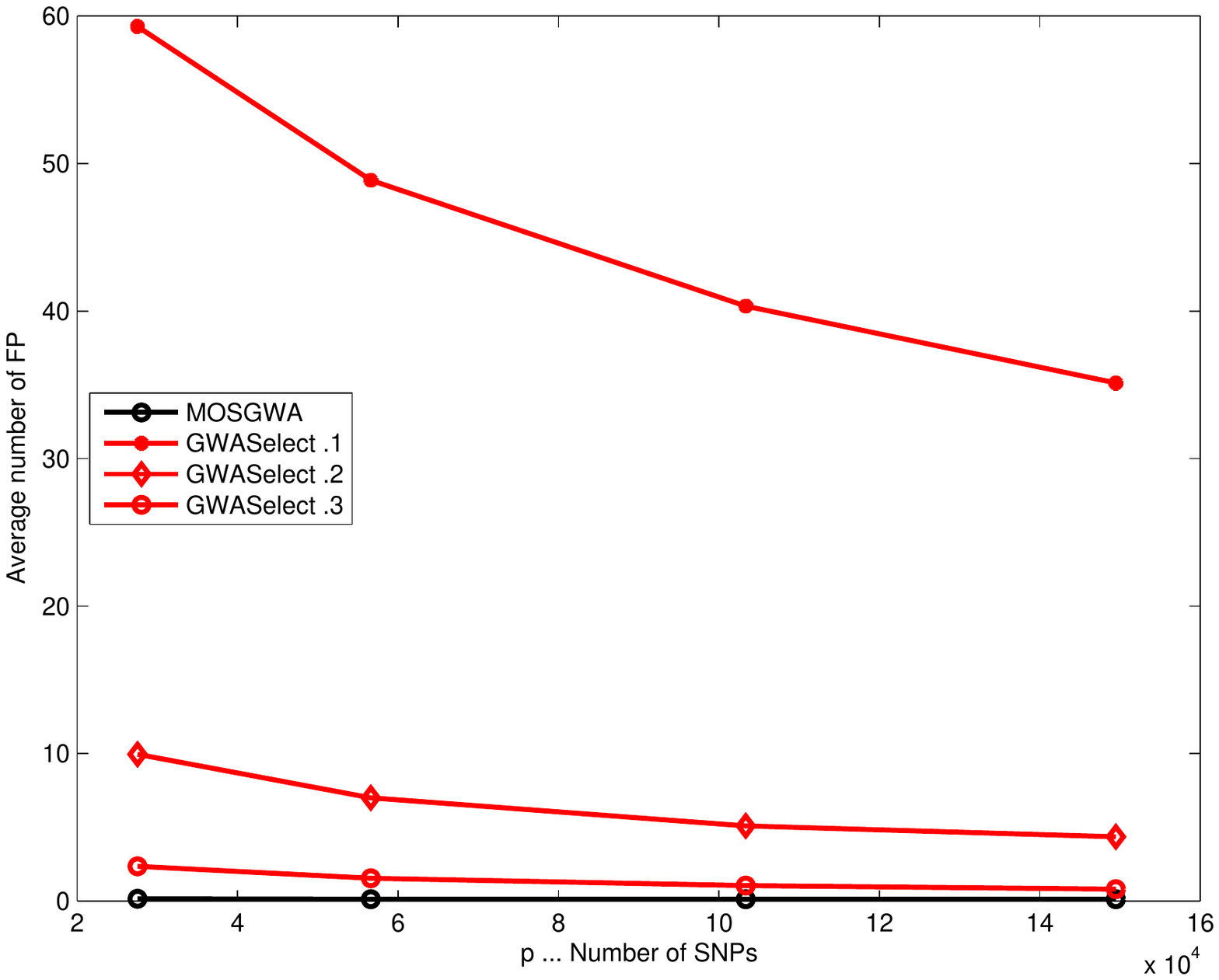}
    \end{center}
  \end{minipage}
 \end{figure}

Our first set of simulations is concerned with controlling the type I error under the global null hypothesis. Simulations are based on real SNP data from $n = 4077$ individuals from the POPRES sample \citep{Nel8}, which are included in the POPRES\_Genotypes\_QC2 dataset. Individuals are randomly allocated as cases and as controls with equal probability; then MOSGWA, HLASSO and GWASelect are applied to evaluate their ability to control the type I error rate. The random allocation was repeated 200 times, and Table \ref{Tab:Null} presents the average number of observed false positives to estimate the per-family error rate. A graphic illustration is provided in Figure \ref{fig:total_null}.

To study the influence of the number of SNPs we performed four different simulations using SNPs only from chromosome 1, from the first two, the first four and the first six chromosomes, respectively. The resulting number of SNPs, $p$, for these four scenarios is given in the first column of Table \ref{Tab:Null}. 
GWASelect and single marker tests tend to report a large number of correlated SNPs, and thus for these two procedures the number of false positives was obtained by counting clusters of neighboring SNPs as described by \cite{He11}, where clustering was performed with the algorithm of \cite{From}. For HLASSO and MOSGWA no such clustering is necessary, because these algorithms tend to select only one representative for a genomic region anyway.

\begin{table}[t]
\caption{Mean number of false positives under the global null
hypothesis. $p$ refers to the total number of SNPs. The methods analyzed were MOSGWA (MOS), HLASSO with three different choices of the parameter $\alpha$, GWASelect with three different choices of the stable-selection-threshold $\xi$, and single marker tests (SM) with Benjamini Hochberg procedure at level $\alpha = 0.05$.} \label{Tab:Null} 
\begin{center}
\begin{tabular}{l|cccccccc}
$p$ & MOS & \multicolumn{3}{c}{HLASSO ($\alpha$)}&
\multicolumn{3}{c}{GWASelect ($\xi$)} & SM 
\\
        & & $.3/p$ & $.2/p$ & $.1/p$ & .1 & .2& .3 &   \\
\noalign{\smallskip}\hline\noalign{\smallskip} 
27520  &   0.16 & 0.22 & 0.12 & 0.03 & 51.1 &  9.34 & 2.34 & 0.04 \\
56629 &   0.13 & 0.13 & 0.07 & 0.03 & 43.34 &  6.73 & 1.51 & 0.04 \\
103348 &   0.13 & 0.18 & 0.11 & 0.04 & 36.64 &  4.90 & 1.06& 0.04 \\
149478 &   0.13 & 0.16 & 0.05 & 0.03 & 32.22 &  4.19 & 0.79& 0.06  
\end{tabular}
\end{center}
\end{table}

Table \ref{Tab:Null} illustrates that MOSGWA controls the type I error under the total null hypothesis irrespective of the number of SNPs. For MOSGWA no parameter needs to be tuned, whereas the latest version of HLASSO allows to choose a parameter $\alpha$ which corresponds to an uncorrected nominal significance level. We consider three Bonferroni-like choices of the form $\alpha \in \{0.3/p,0.2/p,0.1/p\}$. The results from Table \ref{Tab:Null} indicate that the type I error tends to remain below the nominal level, which is not too surprising given the positive correlations between neighboring SNPs due to linkage disequilibrium. Type I error rates for  $\alpha = 0.3/p$ are closest to those from MOSGWA, and therefore in Section \ref{CompTrait} HLASSO will be  used with this parameter setting. 

The size of the model selected by GWASelect depends on  the stable-selection-threshold $\xi$. \cite{He11}  recommend a choice of $\xi$ between 0.1 and 0.2, but the results from Table \ref{Tab:Null} show that for these parameter settings GWASelect completely fails to control the type I error rate under the global null. We therefore considered additionally $\xi = 0.3$, for which the per-family error rate is controlled at least to some extent. 
Interestingly the type I error from GWASelect decreases when the number of SNPs increases.

The last column of Table \ref{Tab:Null} provides the results of single marker tests performed with PLINK \citep{PLINK}. We considered  logistic regression models for each marker including the first four principle components of all SNP genotypes as covariates to account for population structure. This kind of adjustment is in principle not necessary when simulating under the total null hypothesis, but it becomes important for the simulations of Section \ref{CompTrait}. We applied the Benjamini Hochberg procedure to account for multiple testing, and we can see that under the total null hypothesis the single marker tests control the type I error rate pretty much at the nominal level $\alpha = 0.05$.

\subsection{Complex trait} \label{CompTrait}

The second set of simulations is concerned with the power to detect causal SNPs. To this end 
we consider again the $149478$ SNPs from Chromosome 1-6 for the 4077 individuals from the POPRES sample. Simulations are performed for three scenarios, which include 6, 12, and 24 causal SNPs, respectively. All causal SNPs are common (MAF $> 0.3$), equally distributed over the six chromosomes, and with pairwise correlation $\rho < 0.1$ for each pair.  Disease risk was computed for each individual according to equation (\ref{logreg}), based on which for each scenario 200 case-control data sets were sampled. Effect sizes $\beta_j$ were ranging in the interval $[0.2, 0.28]$, yielding causal SNPs with intermediate power. The coefficient of the intercept $\beta_0$ was chosen such that the number of cases and controls in each simulation run was more or less identical.

Before analyzing the data in each scenario half of the causal SNPs were removed, mimicking the situation where SNPs associated with a trait are not causal themselves, but only in linkage disequilibrium with the cause. SNPs to be removed were selected in such a way that there actually were SNPs in linkage disequilibrium, to make it possible for these signals to be detected.  The simulated data were then analyzed with MOSGWA, HLASSO (using parameter $\alpha = 0.3/p$), GWASelect using parameters $\xi \in \{0.1, 0.2, 0.3\}$ and with single marker tests (as previously including the four leading principal components in logistic regression models and applying Benjamini Hochberg procedure at nominal level $\alpha = 0.05$). 

\begin{figure}[!t]
\caption{Simulation results under an alternative with $k = 6$ causal SNPs. 
} \label{SFig:k6}
 \ \\
  
\begin{minipage}{5cm}  Power \\[6mm]  \centerline{\includegraphics[width = 1.7cm, bb = 200 200 350 550]{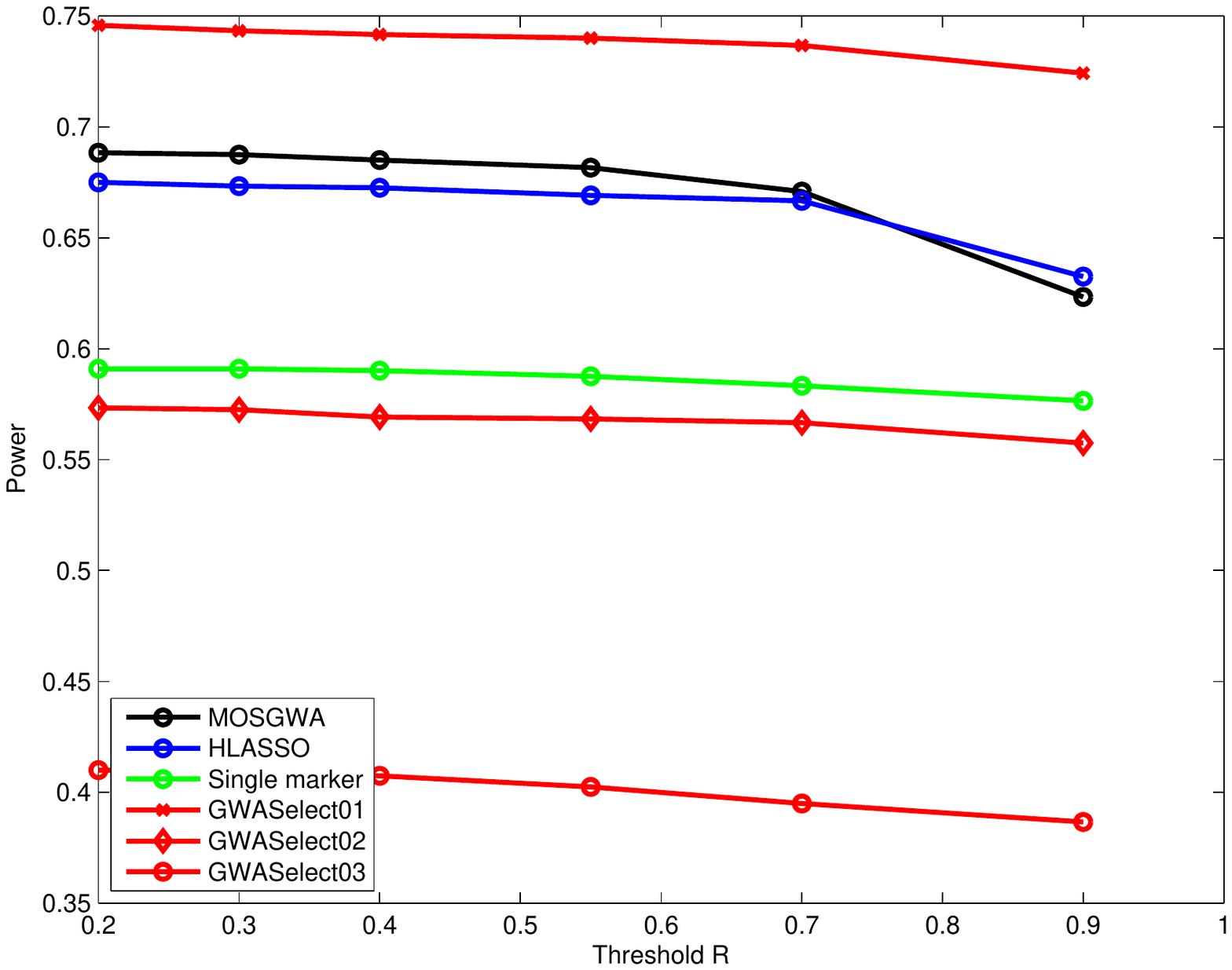}}
\end{minipage}
\hspace{1.4cm}
\begin{minipage}{5cm} Average number of FP \\[6mm]  \centerline{\includegraphics[width = 1.7cm, bb = 200 200 350 550]{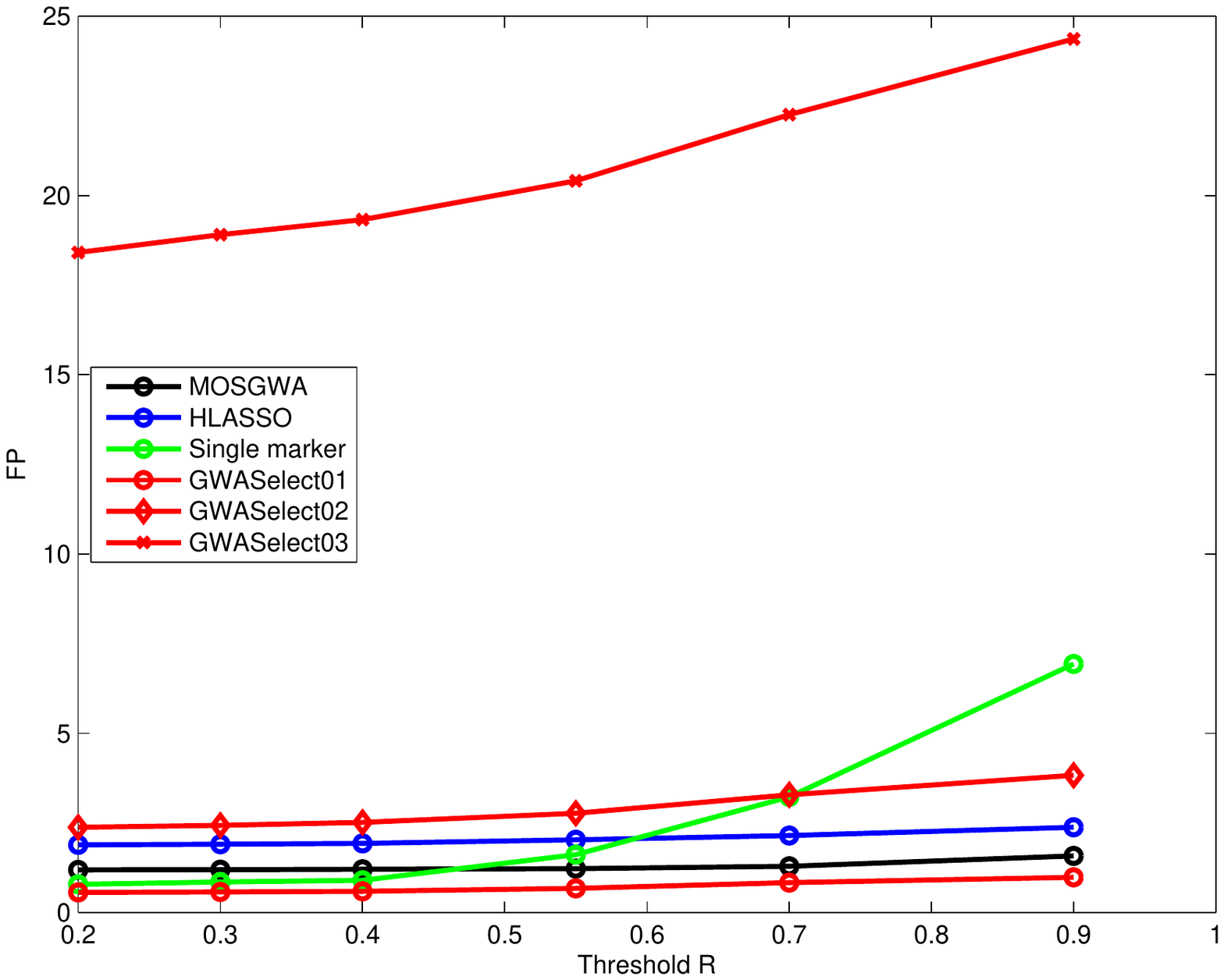}}
\end{minipage}

\ \\

\begin{minipage}{5cm}  Misclassification  \\[6mm]   \centerline{\includegraphics[width = 1.7cm, bb = 200 200 350 550]{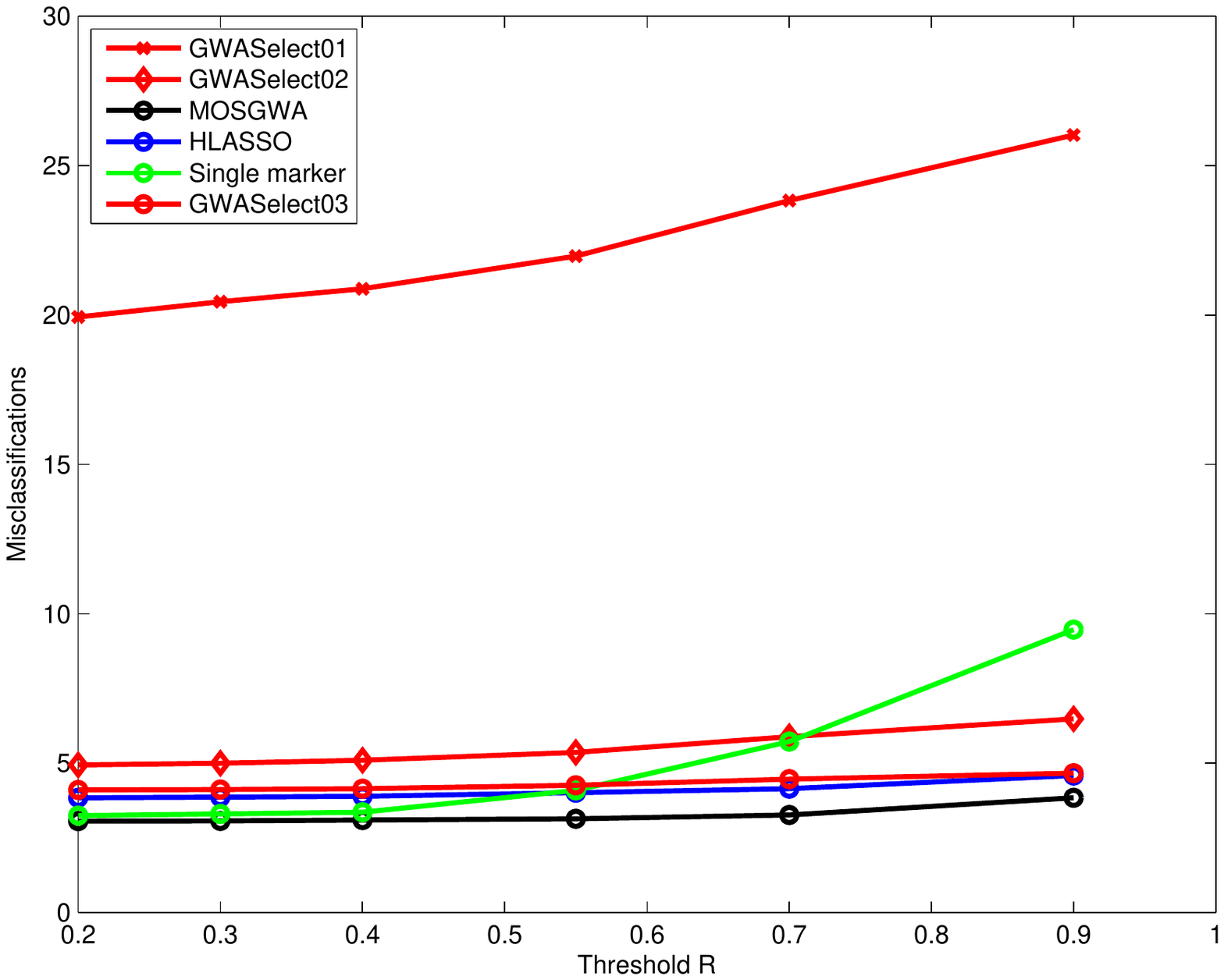}}
\end{minipage}
\hspace{1.4cm}
\begin{minipage}{5cm} FDR   \\[6mm]  \centerline{\includegraphics[width = 1.7cm, bb = 200 200 350 550]{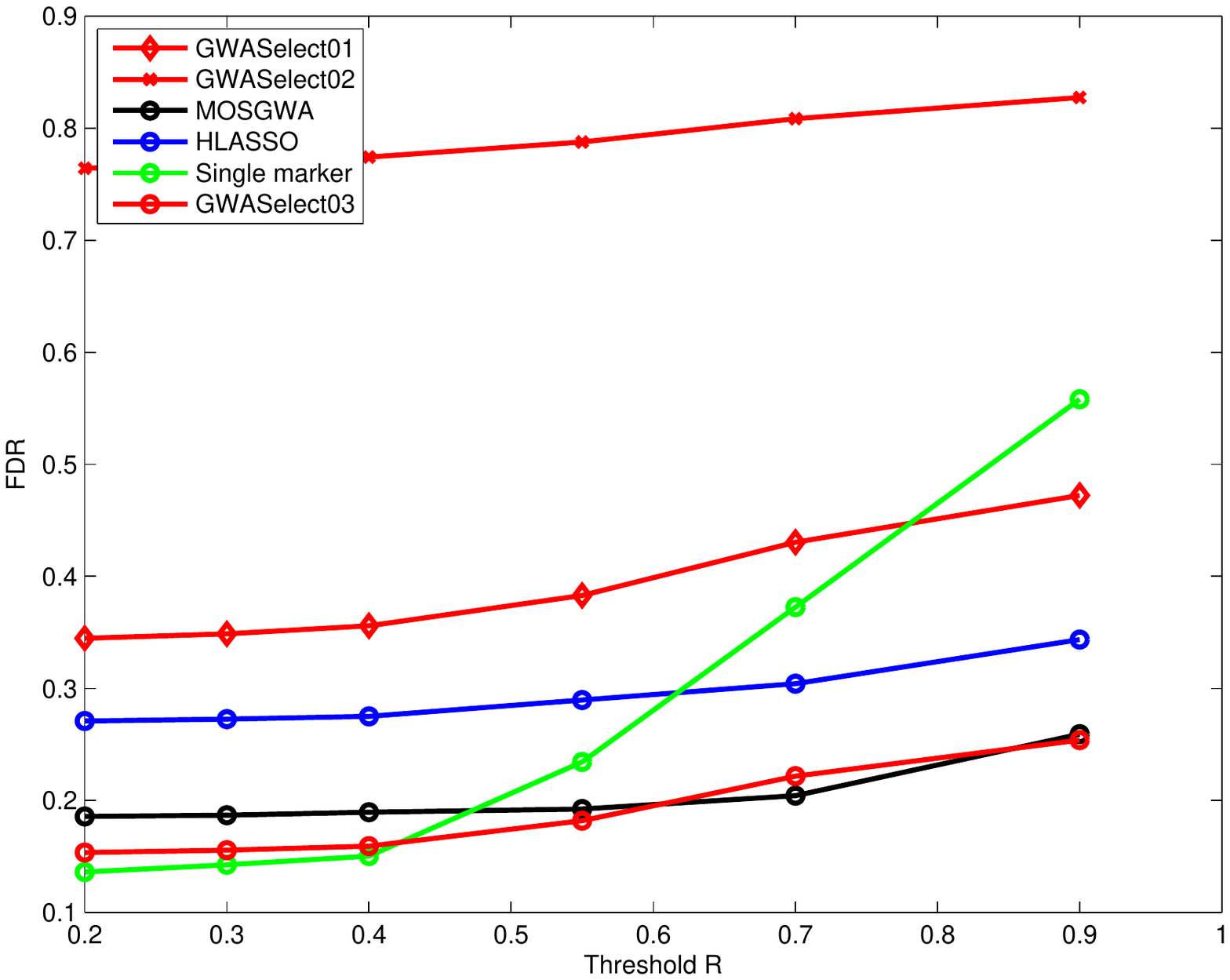}}
\end{minipage}

 \end{figure}

\begin{figure}[!t]
\caption{Simulation results under an alternative with $k = 12$ causal SNPs. 
}  \label{SFig:k12}
 \ \\
  
\begin{minipage}{5cm}  Power \\[6mm]  \centerline{\includegraphics[width = 1.7cm, bb = 200 200 350 550]{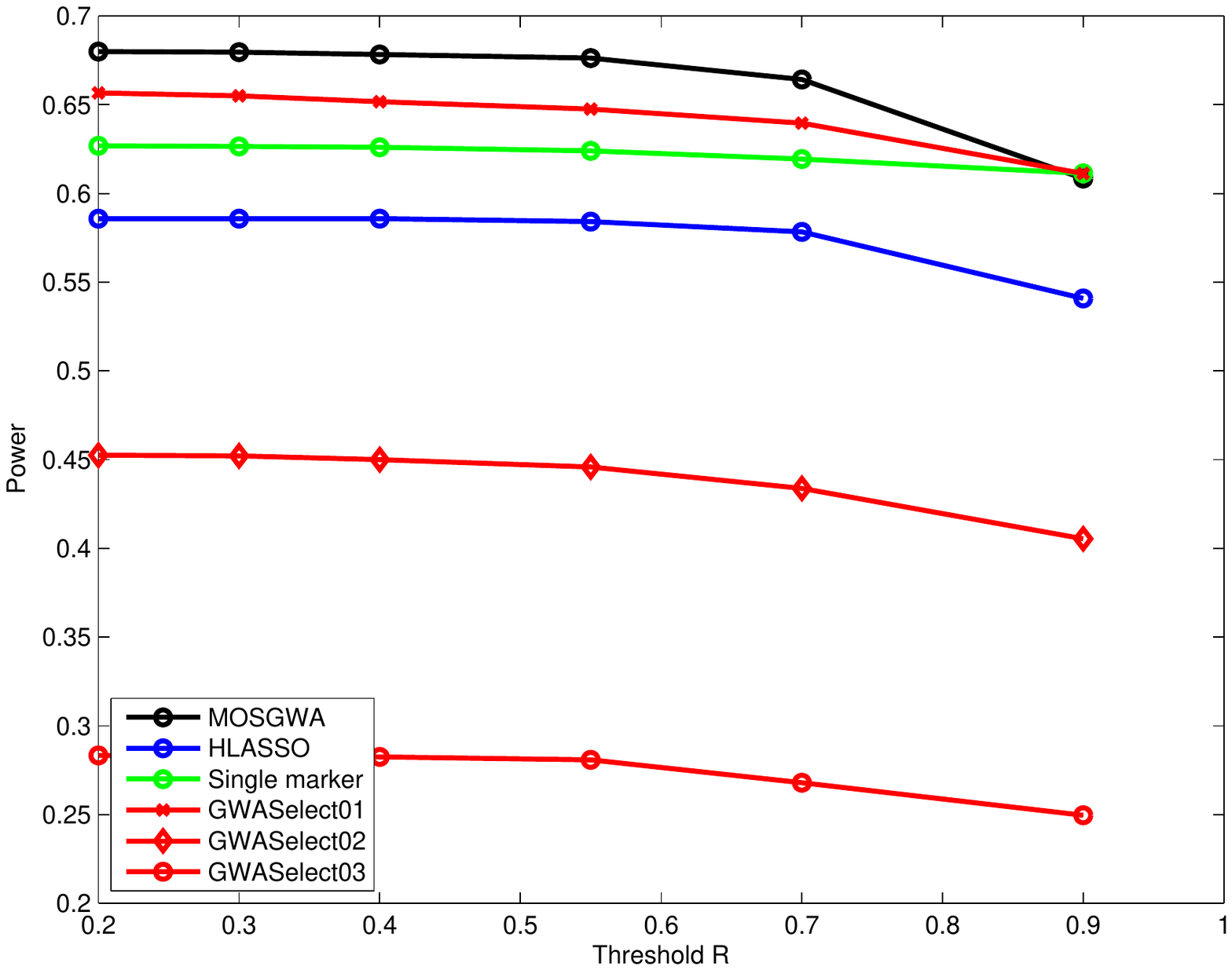}}
\end{minipage}
\hspace{1.4cm}
\begin{minipage}{5cm} Average number of FP \\[6mm]  \centerline{\includegraphics[width = 1.7cm, bb = 200 200 350 550]{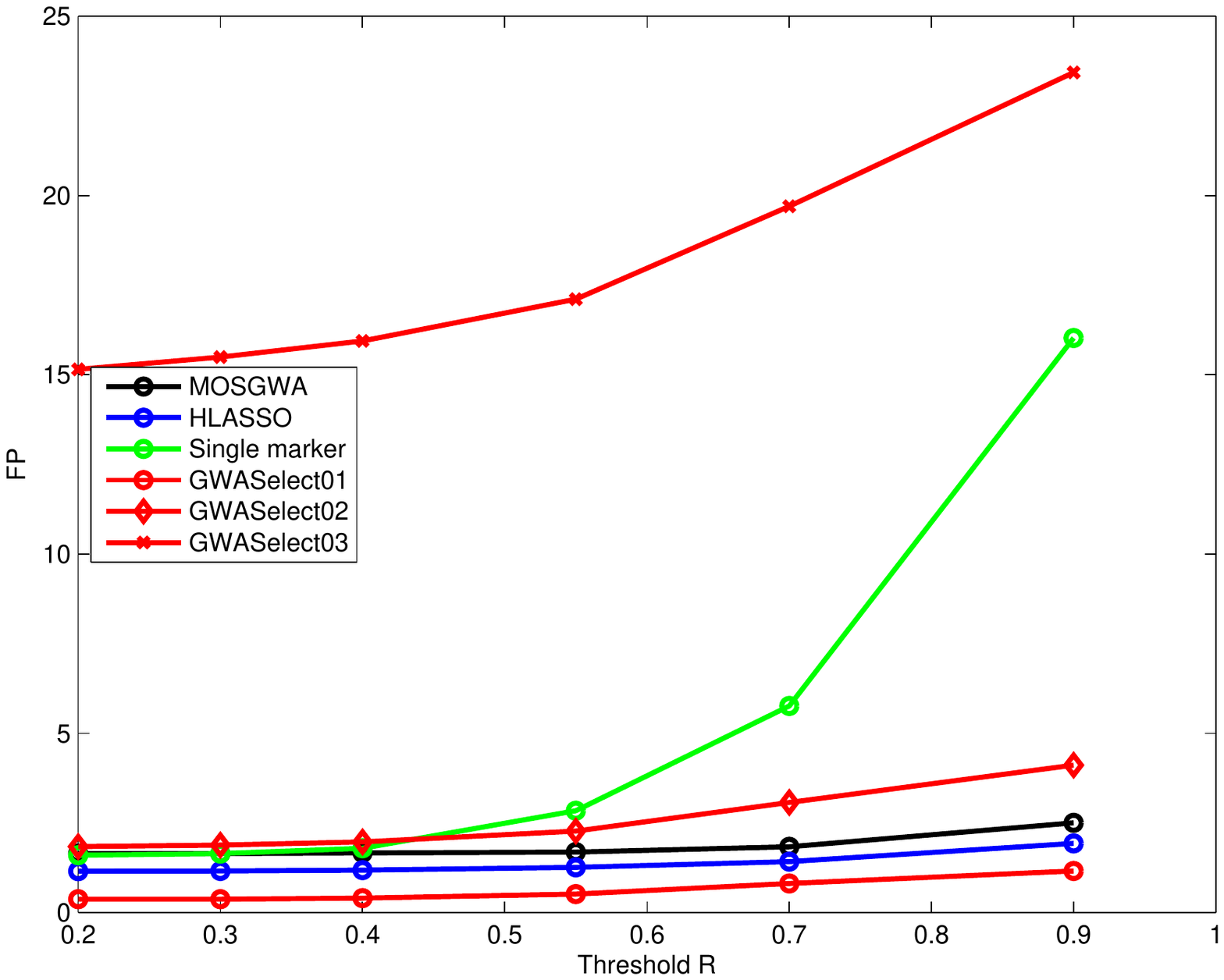}}
\end{minipage}

\ \\

\begin{minipage}{5cm}  Misclassification  \\[6mm]   \centerline{\includegraphics[width = 1.7cm, bb = 200 200 350 550]{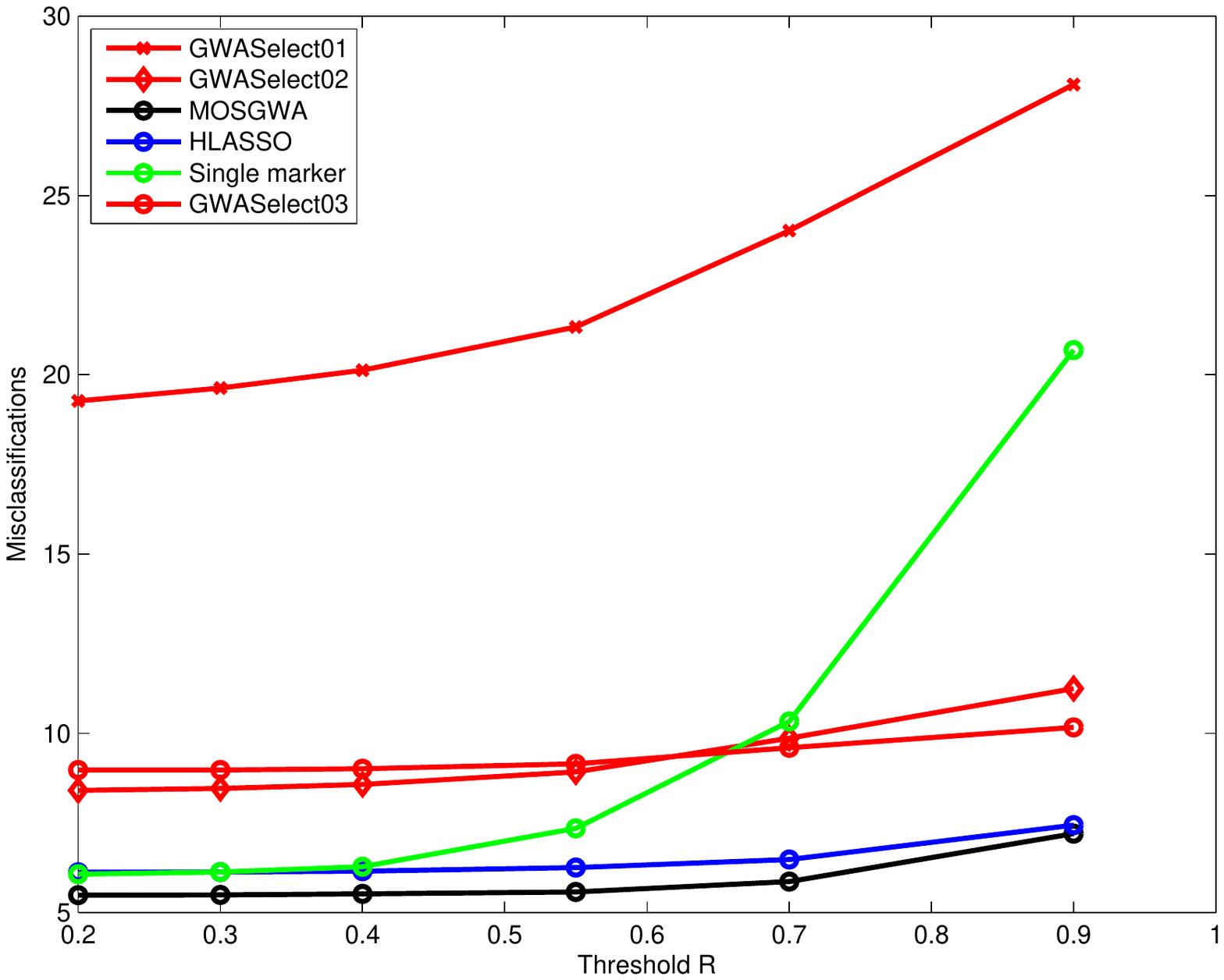}}
\end{minipage}
\hspace{1.4cm}
\begin{minipage}{5cm} FDR   \\[6mm]  \centerline{\includegraphics[width = 1.7cm, bb = 200 200 350 550]{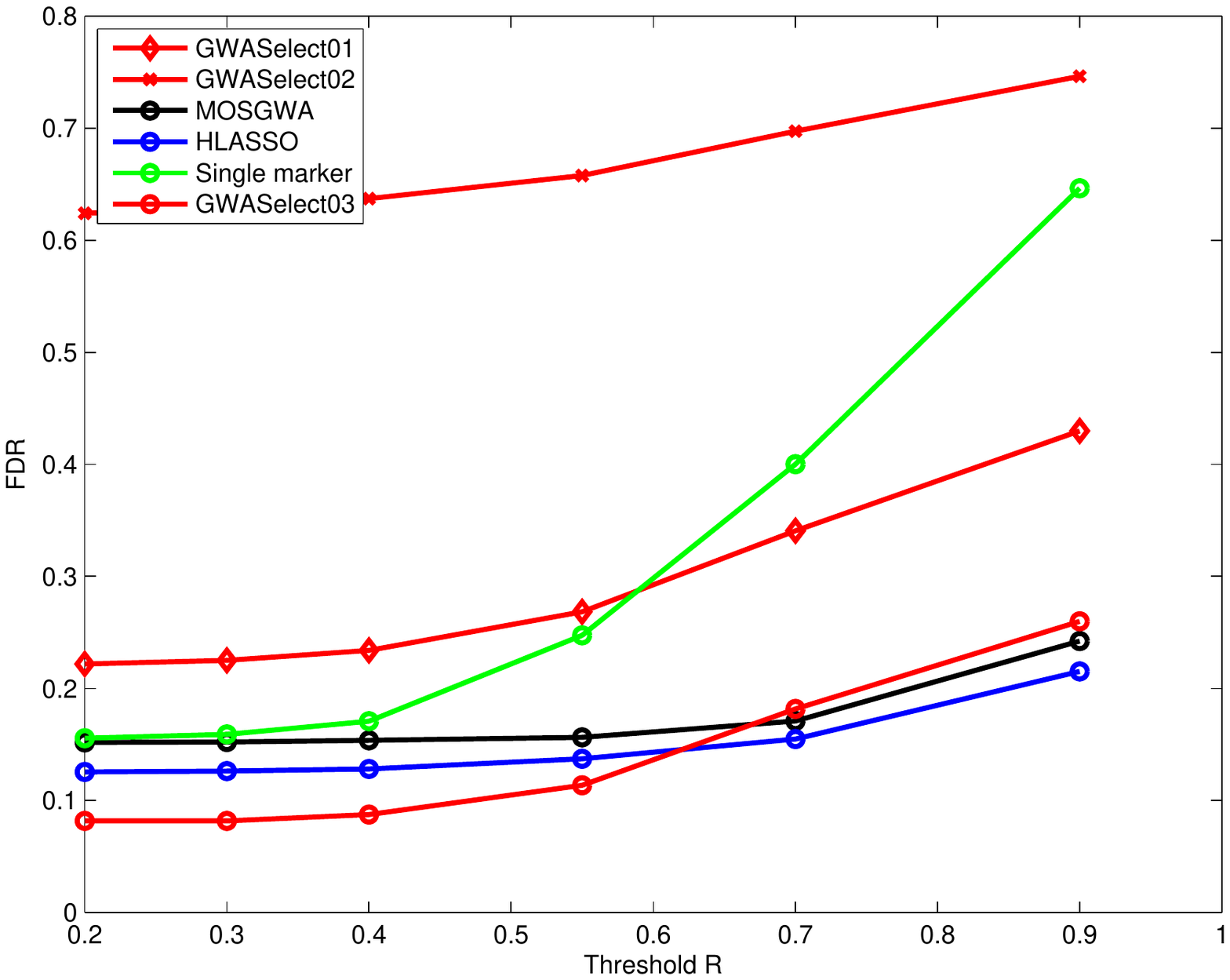}}
\end{minipage}

 \end{figure}

\begin{figure}[!t]
\caption{Simulation results under an alternative with $k = 24$ causal SNPs. 
} \label{SFig:k24}
 \ \\
  
\begin{minipage}{5cm}  Power \\[6mm]  \centerline{\includegraphics[width = 1.7cm, bb = 200 200 350 550]{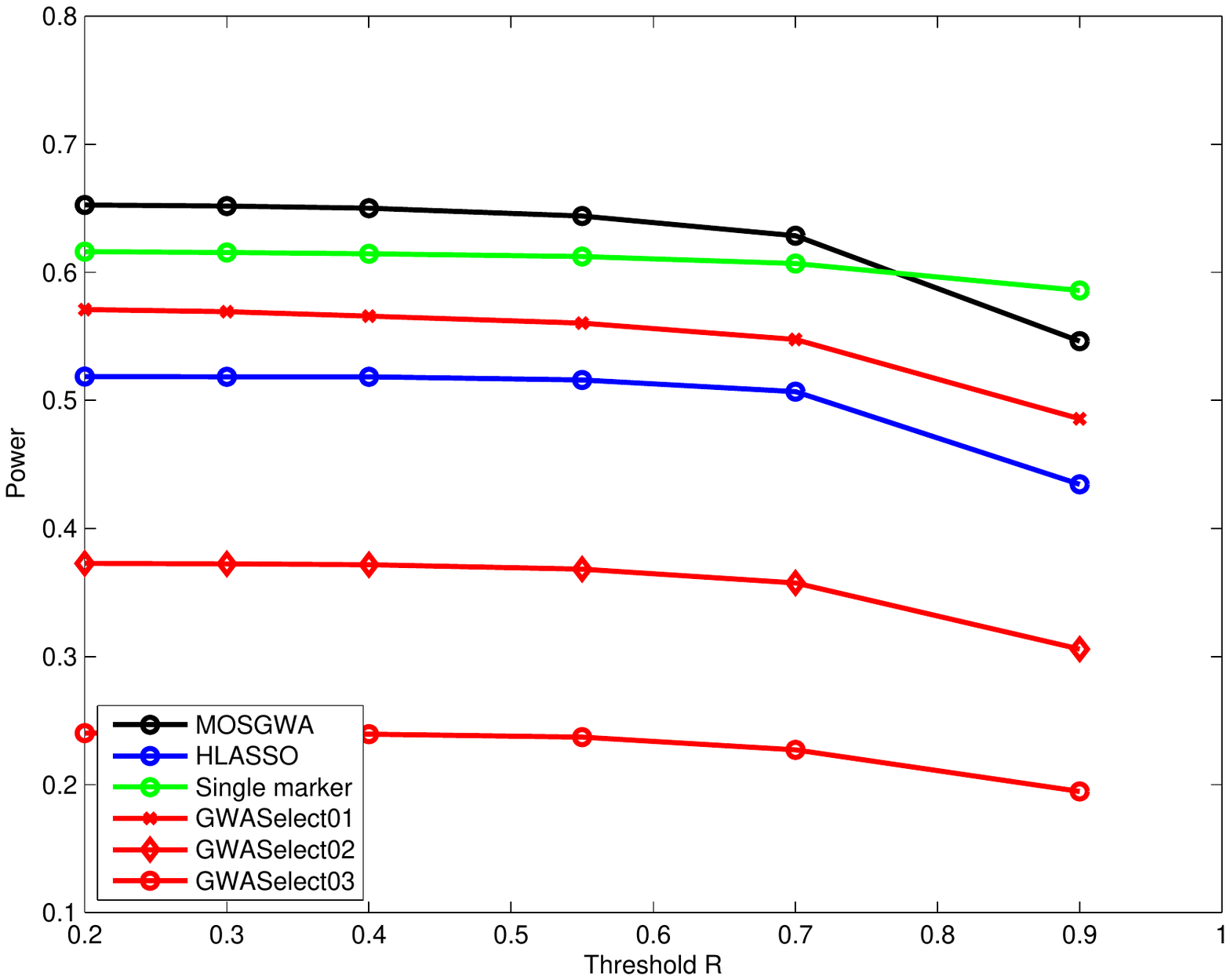}}
\end{minipage}
\hspace{1.4cm}
\begin{minipage}{5cm} Average number of FP \\[6mm]  \centerline{\includegraphics[width = 1.7cm, bb = 200 200 350 550]{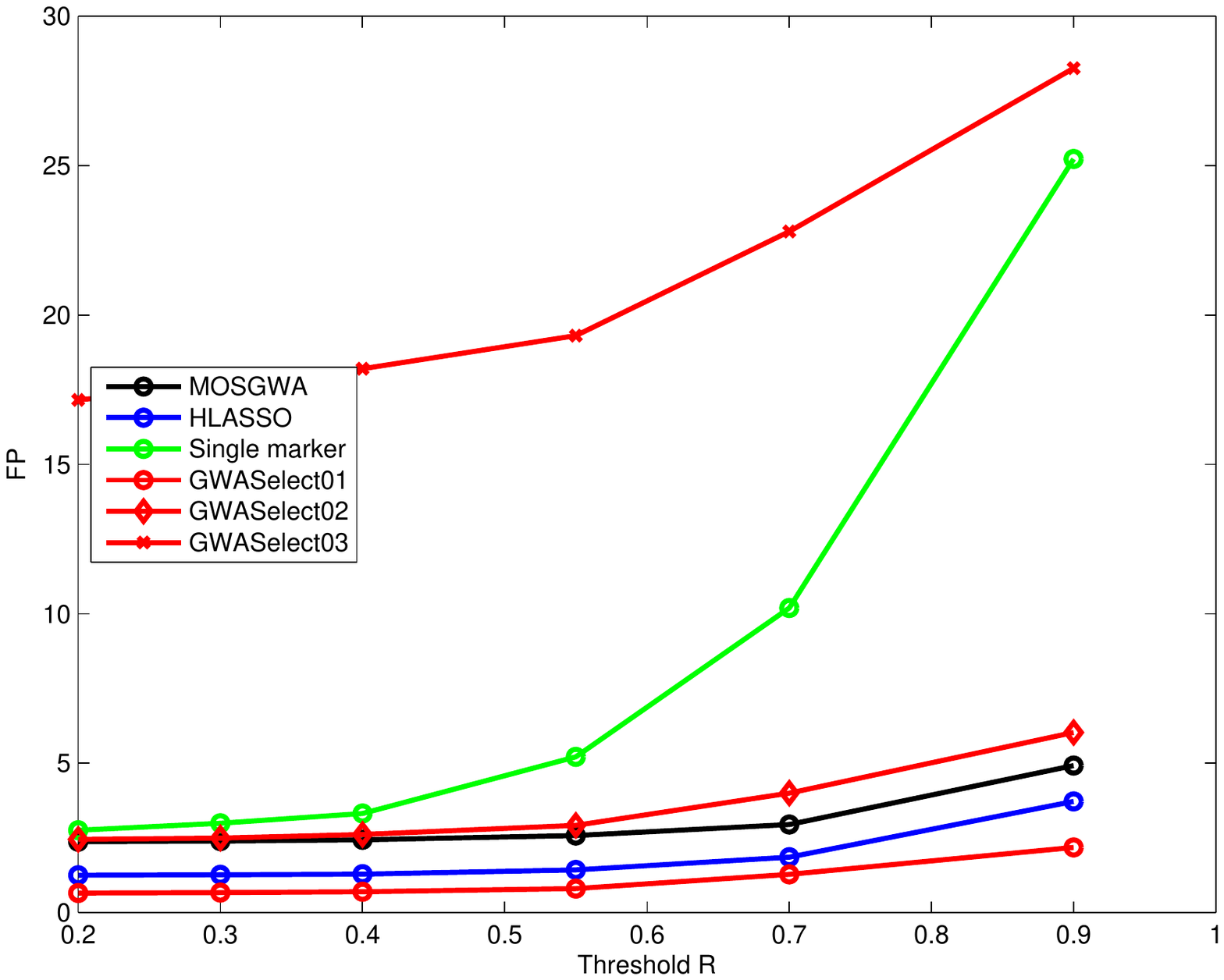}}
\end{minipage}

\ \\

\begin{minipage}{5cm}  Misclassification  \\[6mm]   \centerline{\includegraphics[width = 1.7cm, bb = 200 200 350 550]{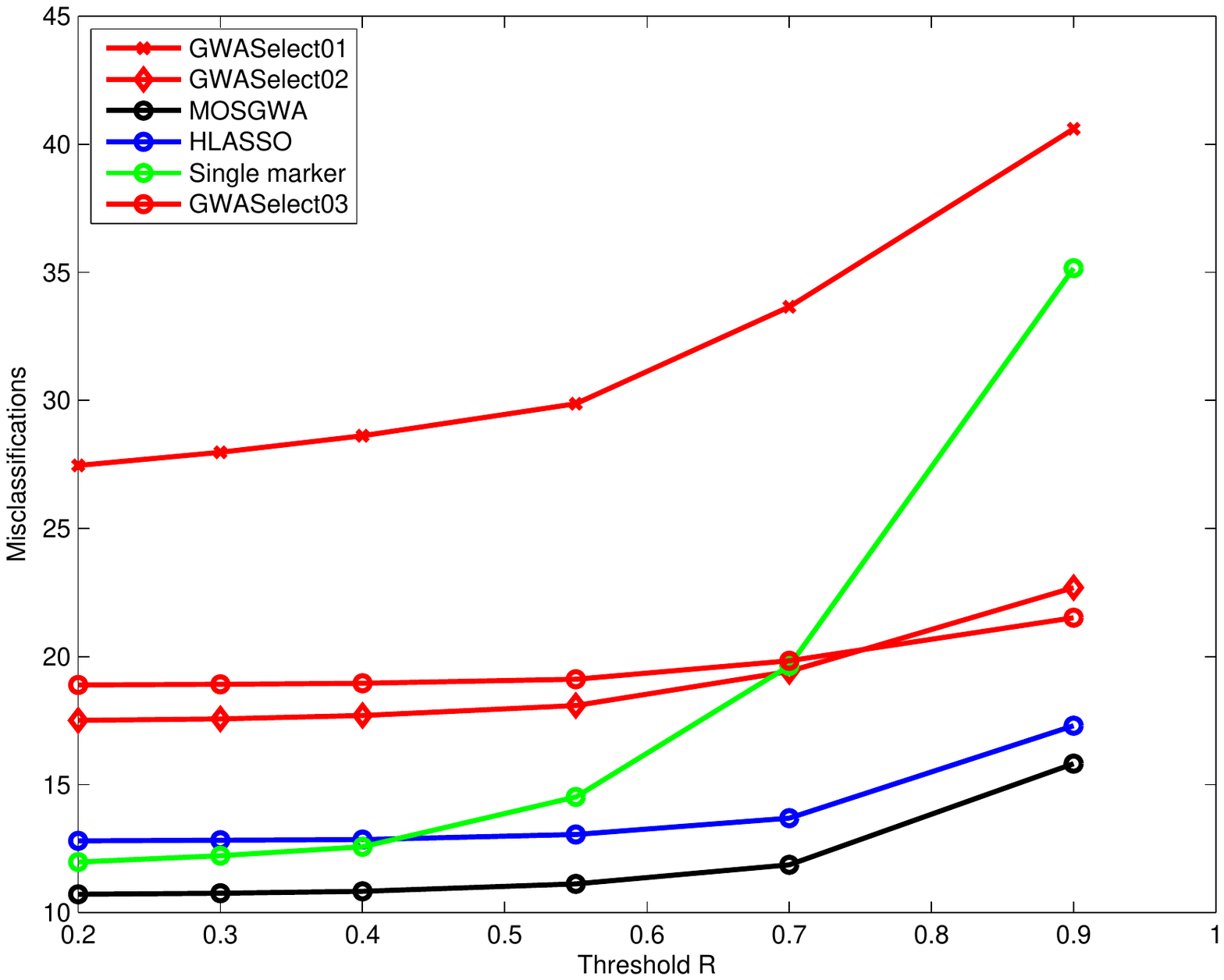}}
\end{minipage}
\hspace{1.4cm}
\begin{minipage}{5cm} FDR   \\[6mm]  \centerline{\includegraphics[width = 1.7cm, bb = 200 200 350 550]{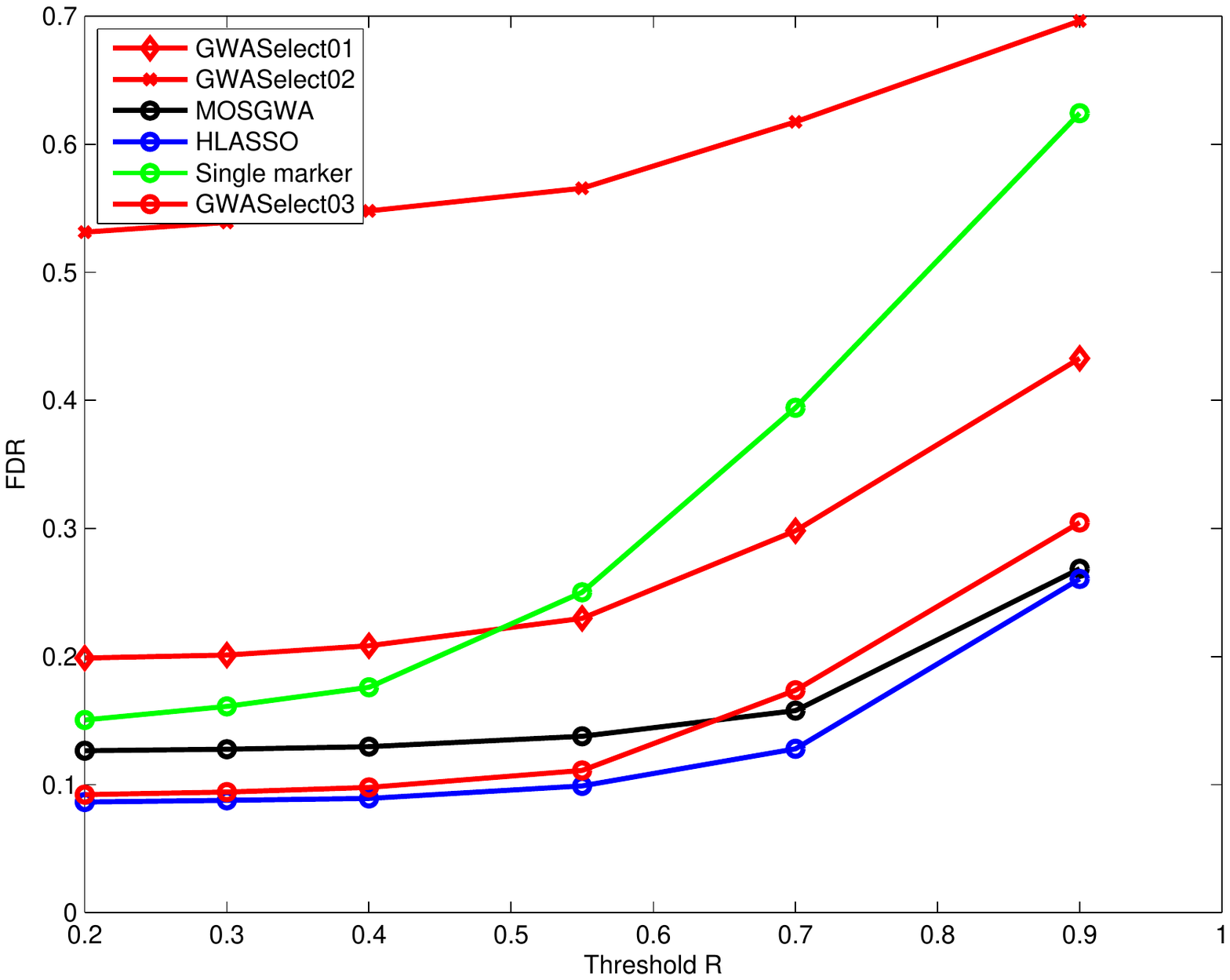}}
\end{minipage}

 \end{figure}

Table \ref{Tab:Complex} and Figures \ref{SFig:k6},  \ref{SFig:k12} and \ref{SFig:k24} summarize the corresponding results in terms of estimated power (which is just the average number of correctly detected signals divided by the total number of causal SNPs), the average number of false positives, the average number of misclassifications (that is false positive plus missed causal SNPs) and the estimated false discovery rate. 


%
%

%
%

A crucial point in computing all these statistics is the definition of true positives and false positives. Of course we know the  SNPs which we used to simulate the data, which we will call causal SNPs or correct SNPs. Now there might be several SNPs in close linkage disequilibrium with a causal SNP. Do we count a detected SNP which is strongly correlated with the correct SNP as true positive or as false positive? Furthermore half of the causal SNPs under which we simulated were selected to be removed before analyzing the data. To get reasonable results we thus actually have to count detections which are strongly correlated with a causal SNP as true positives.

%
%

Just like \cite{Hog8} and \cite{He11} we use threshold values $C$ on the correlation between causal SNPs and any detected SNP to determine whether a detected SNP is a true positive. If several detected SNPs are closely correlated to one causal SNP we count all of them as one true positive. As described in the main manuscript we have additionally  clustered the false positives of GWASelect and of the single marker analysis. This was performed with the algorithm described by \cite{From} which computes so called $C$-clusters, that is clusters of SNPs where it is guaranteed that within each cluster all SNPs have pairwise correlation larger than $C$. We used for clustering and for determining true positives always the same constant $C$. Counting the number of clusters rather than the total number of false positives works in favor of the performance of GWASelect and single marker tests. Both for MOSGWA and HLASSO such clustering appears to be unnecessary, because for a genomic region of closely related SNPs usually these procedures select only one representative.


In Figures \ref{SFig:k6},  \ref{SFig:k12} and \ref{SFig:k24}
we illustrate the dependence of the different statistics on the threshold values $C$ for correlations which specify true positives via $|R| > C$. Results were computed for $C \in \{0.2, 0.3, 0.4, 0.55, 0.7, 0.9\}$. One can see that within the range of $0.2 \leq C \leq 0.55$ the dependence on the threshold is relatively minor for all methods. In general the choice of $C$ has the biggest impact on the results of single marker tests. For those the number of false positives grows much faster than for the other methods, because the number of false positive clusters increases with stricter clustering threshold $C$.    Table \ref{Tab:Complex} specifically reports the results for a threshold $C = 0.3$.

\begin{table}[b!]
\caption{Summary of simulation results for complex traits. The average over 200 simulation runs is reported for the number of detected associations (Size), the estimated power, the number of false positive detections (FP), the estimated false discovery rate (FDR) and the average number of misclassifications (Mis). GWASelect performed with parameters $\xi \in \{0.1, 0.2, 0.3\}$ is abbreviated as  GS $\xi$, MOSGWA as MOS , HLASSO as HL, and single marker tests as SM. } \label{Tab:Complex}
\begin{center}
\begin{tabular}{lcccccc}
& MOS & HL & GS 0.1 & GS 0.2 & GS 0.3 & SM\\ 
\noalign{\smallskip}\hline\noalign{\smallskip} 
\multicolumn{3}{l}{\underline{Scenario 1: ($k = 6$)}}& \multicolumn{3}{l}{} \\
Size &  5.32  &  5.94 &   23.36 &    5.87 &   3.03 & 4.40 \\
Power&  0.69  &    0.67 &     0.74 &     0.57 &    0.41  & 0.59\\
FP   &  1.19 &   1.90 &   18.90 &   2.43 &   0.57  &  0.85\\
FDR  &  0.19 &     0.27&    0.77 &    0.35 &    0.16 &    0.14\\
Mis  &  3.07 &    3.86 &   20.44 & 5.00 &   4.12   & 3.31\\
\noalign{\smallskip}\hline\noalign{\smallskip} 
\multicolumn{3}{l}{\underline{Scenario 2: ($k = 12$)}}& \multicolumn{3}{l}{} \\
Size & 9.80 &    8.19 & 23.35 &   7.31 &   3.77 & 9.17\\
Power& 0.68  &  0.59  &  0.65   &  0.45   &  0.28  &  0.63\\
FP   &  1.65 &    1.16 &   15.49 &    1.88 &    0.37 & 1.65\\
FDR  & 0.15 &    0.17 &    0.63 &    0.23 &     0.08 &    0.16 \\
Mis  &  5.49 &   6.13 &  19.63 &   8.46 &   8.97 &  6.13\\
\noalign{\smallskip}\hline\noalign{\smallskip} 
\multicolumn{3}{l}{\underline{Scenario 3: ($k = 24$)}}& \multicolumn{3}{l}{} \\
Size & 18.03 &   13.71 &   31.29 &   11.43 &    6.43 &   17.76\\
Power& 0.65  &  0.52  &  0.57 &   0.37  &  0.24   & 0.62\\
FP   &  2.39 &   1.27 &  17.63 &   2.49  &  0.67  &  2.99\\
FDR  &  0.13  &  0.09  &  0.54  &  0.20  &  0.09  &  0.16\\
Mis  & 10.75  & 12.83 &  27.97  & 17.56 &  18.91  & 12.22
\end{tabular}
\end{center}
\end{table}

For all three scenarios MOSGWA has the lowest number of misclassifications, which is in accordance with the theoretical results from \cite{FBMC}. Of particular interest is the comparison between MOSGWA and HLASSO. For $k = 6$ MOSGWA has slightly larger power and lower Type I error. For $k = 12$ and specifically for $k = 24$ MOSGWA has much larger power than HLASSO, but also larger Type I error. When the number of causal SNPs is increasing then HLASSO is getting more conservative in comparison with MOSGWA, which is in accordance with the way both algorithms are designed. HLASSO tries to control the FWER at a certain level, whereas MOSGWA tries to control the FDR. According to the theory of \cite{BCFG} and \cite{FB13}, when aiming at a minimal number of misclassifications in a sparse setting it is preferable to control FDR rather than FWER.

Concerning GWASelect just like under the total null hypothesis the choice of $\xi = 0.1$ gives way too large models. The choice of $\xi = 0.2$ works slightly better in terms of controlling the Type I error, but is less powerful than MOSGWA and HLASSO. Given the simulation results under the total null hypothesis one actually should use the setting  $\xi = 0.3$, but then GWASelect is no longer competitive at all in terms of power. 

The single marker tests are performing surprisingly well, and the disadvantage compared to  the model selection approaches is much less than it was observed by \cite{FRTB} for quantitative traits. However, we believe that this is mainly due to the fact that we have added the first four principle components of SNP genotypes. Although we did not specifically simulate scenarios where population structure would play a major role in itself, when testing a specific SNP the principle components in the model help to adjust for the net effect of all the other causal SNPs. Without adding principle components the performance of the single marker tests was extremely poor. Also note that for MOSGWA we did not specifically take any measures to take into account population structure. Still it is much more powerful than single marker tests in all three scenarios, while  at the same time controlling FDR at a comparable level.

\section{Real Data}
\label{Sec:RealData}

The Wellcome Trust data for genome-wide association studies on seven different diseases \citep{WTCCC} have become benchmark data sets for comparing different algorithms to analyze GWAS.  For each disease approximately 2000 cases were compared with a common set of approximately 3000 controls. More than half a million SNPs were genotyped with the Affymetrix GeneChip 500K, from which less than 400,000 passed quality control. The original analysis from WTCCC was mainly based on single marker tests and identified 24 significant SNPs for the seven diseases. 

The analysis of \cite{He11} using GWASelect resulted in 60 distinct loci. 
Unfortunately we were not able to completely reproduce their results, which has several reasons. First of all \cite{He11} did not fully document their preprocessing of data for quality control, and model selection analysis of GWAS data is extremely sensitive with respect to the set of SNPs being studied. Furthermore they seem to have used not always  the same parameter $\xi$ for d-GWASelect, but adapted this for different diseases. We provide next a detailed description of the preprocessing steps we performed, and how they differ from \cite{He11}. 

\subsection{Data preprocessing of WTCCC data}

We reanalyzed bipolar disorder (BD), coronary heart disease  (CAD),
 hypertension  (HT), Crohn's disease (IBD), rheumatoid athritis (RA), type 1 diabetes (T1D) and type 2 
diabetes (T2D). Like in the original article  \cite{WTCCC} all diseases respectively are compared with the same control group of 3000 individuals. In the following filenames we will denote by $<$DIS$>$ any of the seven abbreviations for diseases given above.

Starting from the already imputed WTCCC data sets we removed individuals for each disease and for the control group
  according to the files exclusion-list-05-02-2007-$<$DIS$>$.txt and  exclusion-list-snps-26\_04\_2007.txt. 
 
  After merging cases with controls we removed SNPs with a minimal allelic frequency smaller than 0.01. Furthermore we tested for Hardy Weinberg equilibrium and SNPs with p-values smaller than 0.0001 were also removed.
Finally we only considered SNPs for which the genotype calling algorithm from the WTCCC confirmed good clustering, where we took that information from the file {\scriptsize WTCCC\_summary\_data/7\_Diseases/$<$DIS$>$/basic/snptest\_$<$DIS$>$\_$<$CHR$>$.txt}, where $<$CHR$>$ is the chromosome number ranging from 1 to 22.
According to private correspondence \cite{He11}  had used for similar purposes the file  wtccc\_$<$DIS$>$\_basic\_chr\_$<$CHR$>$.xml, which must have differed from the file we have used´, but is no longer available from WTCCC. In summary, after preprocessing we will have ended up with a different set of SNPs than \cite{He11}, and according to e-mail correspondence with He it is no longer possible to reconstruct which set of SNPs they had used for their own analysis. Not starting from the same set of SNPs could explain why we were not able to completely reproduce the results reported in \cite{He11}.

\subsection{Summary of results}

Like in the simulation study HLASSO was applied with parameter $\alpha = 0.3$, and for GWASelect we present the results again for $\xi \in \{0.1, 0.2, 0.3\}$, in spite of the fact that we have seen that only $\xi = 0.3$ controls the type I error rate under the total null hypothesis. Table \ref{Tab:WTCC} gives for each disease the number of detected SNPs (and associated regions) obtained from the original WTCCC analysis, MOSGWA, HLASSO, and GWASelect. The full information on the detected SNPs is provided in Section \ref{Sec:WTCCC_Detail}.

\begin{table}[t]
\caption{Number of detected SNPs which are associated to the following seven diseases from WTCCC: Bipolar disorder (BD), coronary artery disease (CAD),  hypertension (HT), Crohn's disease (IBD), rheumatoid arthritis (RA), type 1 diabetes (T1D) and  type 2 diabetes (T2D). WTCCC refers to the regions reported by the original publication \citep{WTCCC} in their Table 3, abbreviations for the other algorithms are just like in Table \ref{Tab:Complex}. In brackets we give the number of DNA regions which are covered by the detected SNPs (see Section 5 below for details). The whole HLA region on chromosome 6 is counted as only one region.} \label{Tab:WTCC}
\begin{center}
\begin{tabular}{l|cllllll}
Disease & WTCCC & MOS & HL & GS 0.3 & GS 0.2 & GS 0.1 \\ 
\noalign{\smallskip}\hline\noalign{\smallskip} 
BD & (1)  & 1 (1)&1 (1) &1 (1) &9 (9) &43 (39)\\
CAD& (1)  &  2 (2)&3 (2) &3 (2) &4 (2)&29 (21)\\
HT & (0)  &1 (1) &1 (1) &1 (1) &4 (4) &29 (26)\\
IBD& (9)  &17 (16) &12 (8)  &12 (5)  &15 (6) &32 (19)  \\
RA & (3)  &11 (5)  &12(2)  &1(1)  &1(1) &13 (2)\\
T1D& (7)  &25 (11)  &22 (4)  &12 (2)  &20 (2) & 33 (2)\\
T2D& (3)  &2 (2)  &3 (2) &4 (2) &8 (4) &28 (19)\\
\end{tabular}
\end{center}
\end{table}

The first observation is that GWASelect with parameter $\xi = 0.1$ is selecting in 6 out of 7 diseases a much larger number of SNPs than the other methods. Given the results from the simulation study of Section \ref{Sec:Simulations} we are forced to conclude that most of those SNPs might be false positives, and we will not give the detailed results for GS 0.1 in Section 5 except for rheumatoid arthritis, where interestingly GS 0.1 gives quite similar results to MOSGWA and HLASSO (for a detailed explanation see Section \ref{Sec:WTCCC_Detail}). For Crohn's disease and Type I diabetes the results for $\xi = 0.2$ are relatively close to MOSGWA and HLASSO, whereas for all other diseases $\xi = 0.3$ might be the best choice for GWASelect. The general conclusion is that the results of GWASelect heavily depend on the choice of $\xi$, and it is not really possible to know in advance which choice gives reliable results. 

The comparison between MOSGWA and HLASSO is quite interesting. For the four diseases for which only a small number of SNPs was detected (BD, CAD, HT, T2D) MOSGWA finds exactly the same regions as HLASSO, though in two cases one SNP less.  Similarly MOSGWA has a tendency to select less representatives of a region than HLASSO for the remaining more complex traits. This might have to do with the fact that the coefficient estimates of MOSGWA suffer from even less shrinkage than the estimates from HLASSO. \cite{Hog8} thoroughly discussed the fact that due to shrinkage regressors which enter the model explain less than they would without shrinkage, which results in a higher chance of including further correlated SNPs in the model. This was the main reason why HLASSO works with the NEG prior, which results in less shrinkage than the double exponential prior which corresponds to LASSO.

 Now especially for complex traits MOSGWA tends to select more regions of association than HLASSO. This goes along with the fact that HLASSO is designed to control the FWER, whereas MOSGWA controls the FDR. In case of complex traits MOSGWA is therefore bound to find more SNPs than HLASSO, whereas if there are only few signals both methods behave very similarly.

Looking more closely into the results for IBD, RA and T1D, the first observation is that MOSGWA detects SNPs within all regions which were reported as significant by the \cite{WTCCC} according to their standard analysis. The same is not true for HLASSO, which misses out on one region on chromosome 10 for RA, and on another region on chromosome 16 for T1D. 
On the other hand MOSGWA finds exclusively 7 SNPs for IBD, 4 SNPs for RA, and 6 SNPs for T1D, respectively. These SNPs are highlighted in yellow in the corresponding tables of Section 5. For RA and T1D all these extra SNPs are lying outside of the HLA regions. We are not particularly interested here in genetic loci within the HLA region, which have been intensively studied now for several decades, but specifically on SNPs outside the HLA. 
In the next Section we will provide a thorough discussion on the potential relevance of the SNPs which were exclusively found by MOSGWA. 


\section{WTCCC Results in detail} \label{Sec:WTCCC_Detail}

In this section of the appendix we provide detailed results on the SNPs detected by the different methods of analysis for the seven diseases for which GWAS data are available from \cite{WTCCC}. For all tables the first column gives the reference SNP ID number from dbSNP, followed by the chromosome (Chr) and the position (Pos). The column Gene contains information about the closest lying gene according to the databases dbSNP and ImmunoBase: 
\begin{itemize}
\item dbSNP:  \href{http://www.ncbi.nlm.nih.gov/SNP/}{http://www.ncbi.nlm.nih.gov/SNP/} 
\item ImmunoBase:  \href{http://www.immunobase.org/}{http://www.immunobase.org/} 
\end{itemize}

The final columns have bullets whenever a SNP was detected by MOSGWA (M), Hlasso (HL), GWASelect with parameter $\xi = 0.3$ (G3)  or  $\xi = 0.2$ (G2).  According to the simulation results from Section 3 we believe that the large number of additional SNPs detected by GWASelect with $\xi = 0.1$ (G1) will include mainly false positives. Therefore we do not present detailed results for G1, with the exception of rheumatoid arthritis which has a rather particular genetic constellation. If many  neighboring SNPs are reported by different methods we consider such groups of SNPs as genetic regions, and we label such groups in the tables using background colors. The crosses in the last column (W) indicate regions which were reported in \cite{WTCCC}.

\subsection{Bipolar disorder (BD)}

\begin{tabular}{lrrllllll}
 dbSNP &Chr&Pos&Gene&M&HL&G3&G2\\ \hline
rs2953145 &2 &241515596 &RNPEPL1 & & & &\textbullet   \\
rs4627791 &3 &32347824 &CMTM8 & & & &\textbullet  \\
rs715891 &5 &145986083 &PPP2R2B & & & &\textbullet  \\
rs10993706 &9 &93602967 &SYK & & & &\textbullet  \\
rs11622475 &14 &104509076 &TDRD9 & & & &\textbullet  \\
rs2576561 &16 &55470974 &MMP2 & & & &\textbullet  \\
rs7243929 &18 &8455102 &PTPRM & & & &\textbullet  \\
rs12980129 &19 &22908911 &ZNF99 & & & &\textbullet  \\
rs2837588 &21 &41748059 &DSCAM &\textbullet &\textbullet &\textbullet &\textbullet  \\
\end{tabular}\\

SNP rs2837588, which is an intron of DSCAM, was found by all methods, though it was not reported in \cite{WTCCC}. On the other hand the  only SNP reported by WTCCC, rs4202459, was not detected by any of the algorithms we analyzed here. This can be quite easily explained because all algorithms we study here are based on models incorporating linear trends, and the trend p-value of rs4202459 is quite large (2.19 E-04 according to the WTCCC manuscript). That rs2837588 was not reported by WTCCC might have to do with the large number of missing values for this SNP. Results reported  in \cite{WTCCC} are not based on imputed data, and imputation changes the marginal p-value for this SNP considerably. More recent research indicates that there actually might be a connection between bipolar disorder and DSCAM \citep{Amano}, although in  general according to a recent large GWAS \citep{MP13} it appears to be extremely difficult to identify robust and replicable genetic causes for psychiatric disorders. Thus all the other SNPs reported by G2 have a good chance to be false positives.

\subsection{Coronary artery disease (CAD)}

\begin{tabular}{lrrllllllc}
 dbSNP &Chr&Pos&Gene&M&HL&G3&G2&W\\ \hline
rs906766 &3 &150811294 &MED12L &\textbullet &\textbullet &\textbullet &\textbullet & \\
rs10965219 &\cellcolor{blue!25}9 &\cellcolor{blue!25}22053687 &CDKN2B-AS & &\textbullet &\textbullet &\textbullet & \cellcolor{blue!25}x \\
rs6475606 &\cellcolor{blue!25}9 &\cellcolor{blue!25}22081850 &CDKN2B-AS & & & &\textbullet &\cellcolor{blue!25}x \\
rs1333049 &\cellcolor{blue!25}9 &\cellcolor{blue!25}22125503 &DMRTA1 &\textbullet &\textbullet &\textbullet &\textbullet & \cellcolor{blue!25}x \\
\end{tabular}\\

Here the original WTCCC study reported only one region, whereas all methods studied here detect two regions. Note that MOSGWA selects only one representative of the region reported by WTCCC, whereas HLASSO reports two representatives. This is a pattern we will see again several times, for example in type II diabetes. A possible explanation why HLASSO prefers to choose more representatives of a region than MOSGWA might be that MOSGWA is based on model selection criteria which impose less shrinkage on the coefficients than HLASSO does. The effect of shrinkage on the number of selected correlated SNPs is thoroughly described by \cite{Hog8}.

\subsection{Hypertension (HT)}

\begin{tabular}{lrrllllll}
 dbSNP &Chr&Pos&Gene&M&HL&G3&G2\\ \hline
rs7961152 &12 &24981611 &BCAT1 & & & &\textbullet  \\
rs921535 &15 &74111343 &TBC1D21 & & & &\textbullet  \\
rs16945811 &17 &1294614 &YWHAE &\textbullet &\textbullet &\textbullet &\textbullet  \\
rs1022684 &20 &18487506 &SEC23B & & & &\textbullet  \\
\end{tabular}\\

The original WTCCC study did not report any SNP associated with hypertension, whereas all methods studied here 
report rs16945811  on chromosome 17. The reason for this is again that we work with imputed data, for which the marginal p-value of this SNP is considerably smaller than for the original
unimputed data. The other SNPs reported by G2 are again very likely to be false positives.

\subsection{Crohn's disease (IBD)}

\begin{tabular}{lrrlllllll}
dbSNP&Chr&Pos&Gene&M&HL&G3&G2&W\\ \hline
rs17375018 &\cb 1 &\cb 67655147 &IL23R CD25 & & & &\textbullet & \cb x\\
rs11805303 &\cb 1 &\cb 67675516 &IL23R &\textbullet &\textbullet &\textbullet &\textbullet& \cb x \\
rs10489629 &\cb 1 &\cb 67688349 &IL23R & &\textbullet &\textbullet &\textbullet& \cb x \\
rs41396545 &\cb 1 &\cb 67689608 &IL23R & & &\textbullet &\textbullet& \cb x \\
rs12119179 &\cb 1 &\cb 67747415 &IL12RB2 & & &\textbullet &\textbullet& \cb x \\
rs12035082 &1 &172898377 &TNFSF18 &\cy \textbullet & & &  &\\
rs10210302 &2 &234158839 &ATG16L1 &\textbullet &\textbullet &\textbullet &\textbullet & x \\
rs11718165 &3 &49696797 &BSN & \textbullet & & & & x \\
rs7726744  &\cb 5 &\cb 40343276 &PTGER4 & & & &\textbullet &  \cb x\\
rs12658567 &\cb 5 &\cb 40391932 &PTGER4 &\textbullet & & & &  \cb x\\
rs16869934 &\cb 5 &\cb 40397352 &PTGER4 & &\textbullet & & &  \cb x\\
rs17234657 &\cb 5 &\cb 40401509 &PTGER4 &\textbullet &\textbullet &\textbullet &\textbullet &  \cb x\\
rs9292777  &\cb 5 &\cb 40437948 &PTGER4 & &\textbullet &\textbullet &\textbullet &  \cb x\\
rs11957215 &\cb 5 &\cb 40445681 &PTGER4 & & &\textbullet &\textbullet &  \cb x\\
rs1505992  &\cb 5 &\cb 40498577 &PTGER4 & & &\textbullet &\textbullet &  \cb x\\
rs11957134 &\ccr 5 &\ccr 150230950 &ZNF300 & &\textbullet & & & \ccr x\\
rs1000113  &\ccr 5 &\ccr 150240076 &ZNF300 &\textbullet & & & & \ccr x\\
rs9405639 &6 &3419149 &SLC22A23 &\cy\textbullet & & & & \\
rs6908425 &6 &20728731 &CDKAL1 &\cy\textbullet & & & & \\
rs4263839 &9 &117566440 &TNFSF15 &\cy\textbullet & & & & \\
rs10761659 &10 &64445564 &ZNF365 &\textbullet &\textbullet & &\textbullet & x\\
rs10883365 &\cb 10 &\cb 101287764 &NKX2-3 & &\textbullet & & & \cb x \\
rs10883371 &\cb 10 &\cb 101292455 &NKX2-3 &\textbullet & & & & \cb x \\
rs11627513 &14 &97539171 &LOC100129345 &\cy\textbullet & & & & \\
rs2076756 &\cb 16 &\cb 50756881 &NOD2 &\textbullet &\textbullet &\textbullet &\textbullet &  \cb x\\
rs7342715 &\cb 16 &\cb 50787483 &CYLD & &\textbullet &\textbullet &\textbullet & \cb x \\
rs2542151 &18 &12779947 &PTPN2 &\textbullet &\textbullet &\textbullet &\textbullet & x\\
rs41526044 &20 &18800670 &SLC24A3 &\cy\textbullet & & & & \\
rs2836753 &21 &40291187 &PSMG1 &\cy\textbullet & & & & \\
\end{tabular} \\

This is the first disease for which model selection approaches become really interesting, because the trait appears to be a complex one. Note that HLASSO finds one region less (on chromosome 3)  than originally reported in \cite{WTCCC}, whereas MOSGWA finds all regions reported by WTCCC plus seven additional ones which are highlighted in yellow in the table above.  At least five of those have been mentioned meanwhile in the literature on Crohn's disease, which means that MOSGWA would have detected a number of SNPs associated with Crohn's disease which were later confirmed by independent studies.

Let's look at those SNPs in detail. The first SNP rs12035082 is close to rs12037606 which was actually reported in \cite{WTCCC} among the SNPs which showed moderate evidence of association, and was later confirmed to be associated with CD by \cite{WS09}. Similarly rs6908425 was reported as being moderately associated in \cite{WTCCC}, and could later be confirmed in an independent study \citep{BH08}. 
rs4263839 was not reported by WTCCC, but it was among the list of confirmed SNPs given by \cite{BH08}. Furthermore it has later been shown to be associated with irritable bowel syndrome \citep{Z11}.
rs2836753 is in close linkage disequilibrium with rs2836754, which was related to Crohn's disease in \cite{PB07}.

 rs6908425 is an intron from the CDKALI gene on chromosome 6 and has been confirmed to be associated to Crohn's disease (see for example the evidence provided in  \href{http://snpedia.com/index.php/Rs6908425}{snpedia} for this SNP.  rs9405639 on chromosome 9 lies within the intron of SLC22A23 gene, which is also well known to be related to Crohn's disease, see for example \\ \href{http://www.immunobase.org/page/Overview/display/gene_id/63027}{http://www.immunobase.org/page/Overview/display/gene\_id/63027}. 

The only two SNPs detected by MOSGWA based on the WTCCC data which have not been  confirmed in the literature are rs11627513 and rs41526044. rs11627513 is relatively close to the IL23R which is known to be associated with Crohn's disease \\ (see \href{http://www.immunobase.org/page/Overview/display/gene_id/149233}{http://www.immunobase.org/page/Overview/display/gene\_id/149233}). and it was mentioned in \cite{HLM} as a potentially associated gene.
Up to our knowledge only for rs11627513 nothing is known, and this might well be a false positive.
Remember that MOSGWA is designed to control the FDR approximately at a level of 10\%, and thus one would actually expect 2 false positive SNPs within this model.

\subsection{Rheumatoid arthritis (RA)}


\begin{tabular}{lrrllllllll}
dbSNP&Chr&Pos&Gene&M&HL&G3&G2& G1 & W\\ \hline
rs6679677 &1 &114303808 &RSBN1 &\textbullet &\textbullet & & &\textbullet & x \\
rs3132671 &\cb 6 &\cb 30178287 &TRIM26 &\textbullet &\textbullet & & & &  \cb x\\
rs3099844 &\cb 6 &\cb 31448976 &MICB & &\textbullet & & & & \cb x \\
rs707974 &\cb 6 &\cb 31629499 &BAT4 & &\textbullet & & &\textbullet & \cb x \\
rs2075800 &\cb 6 &\cb 31777946 &HSPA1L & & & & &\textbullet & \cb x \\
rs17421624 &\cb 6 &\cb 32066177 &TNXB & & & & &\textbullet & \cb x \\
rs3134926 &\cb 6 &\cb 32200147 &NOTCH4 & &\textbullet & & &\textbullet & \cb x \\
rs9267954 &\cb 6 &\cb 32213052 &NOTCH4 & & & & &\textbullet & \cb x  \\
rs910049 &\cb 6 &\cb 32315727 &C6orf10 & & & & &\textbullet & \cb x \\
rs9268418 &\cb 6 &\cb 32343686 &C6orf10 &\textbullet &\textbullet & & & & \cb x \\
rs9268560 &\cb 6 &\cb 32389512 &HLA-DRA & &\textbullet & & &\textbullet &  \cb x\\
rs9268853 &\cb 6 &\cb 32429643 &HLA-DRB5 & &\textbullet & & & & \cb x \\
rs9268858 &\cb 6 &\cb 32429758 &HLA-DRB5 &\textbullet & & & & & \cb x \\
rs9272219 &\cb 6 &\cb 32602269 &HLA-DQA1 & & & & &\textbullet & \cb x \\
rs2856688 &\cb 6 &\cb 32654640 &HLA-DQB1 &\textbullet & & & & & \cb x \\
rs7775228 &\cb 6 &\cb 32658079 &HLA-DQB1 & &\textbullet & & &\textbullet &  \cb x\\
rs6457617 &\cb 6 &\cb 32663851 &HLA-DQB1 & &\textbullet &\textbullet &\textbullet &\textbullet & \cb x  \\
rs9275418 &\cb 6 &\cb 32670244 &HLA-DQB1 &\textbullet &\textbullet & & &\textbullet & \cb x \\
rs9275572 &\cb 6 &\cb 32678999 &HLA-DQA2 & & & & &\textbullet & \cb x \\
rs3128963 &\cb 6 &\cb 33055780 &HLA-DPB1 &\textbullet &\textbullet & & & & \cb x \\
rs12536071 &\ccr 7 &\ccr 42428460 &GLI3 &\cy\textbullet & & & & &  \\
rs12531052 &\ccr 7 &\ccr 42428629 &GLI3 &\cy\textbullet & & & & &  \\
rs2104286 &10 &6099045 &IL2RA &\cy\textbullet & & & & &  \\
rs1945076 &11 &106382378 &GUCY1A2 &\cy\textbullet & & & & &  \\
\end{tabular}\\

All SNPs detected on chromosome 6 belong to the so called HLA region which has been well known for a long time to be associated with rheumatoid arthritis \citep{D89}. According to \cite{CV13} HLA genes explain only approximately one-third of the genetic liability of the disease, and a great amount of research has been performed to understand genetic causes beyond HLA. For the rheumatoid arthritis data GWASelect performs very poor,  for parameters $\xi = 0.3$ and $\xi = 0.2$ only one SNP in the HLA region is reported. This is quite easy to understand given the way GWASelect works. In the HLA region there are many highly correlated SNPs, and during the stability selection procedure it is very likely that for different samples different representatives of a cluster of SNPs are chosen. Therefore only for the lowest threshold  $\xi = 0.1$ GWASelect gives results which are more in line with the other methods. This is the reason why \cite{He11} clustered the HLA region before analyzing that data set, but that appears to be quite an extra effort for the user when applying GWASelect. 
Neither MOSGWA nor HLASSO did have particular problems with including the HLA region in the analysis, and both find apart from SNPs in the HLA region also rs6679677 on chromosome 1. This SNP was already reported in \cite{WTCCC}, but according to a recent meta-analysis by \cite{CV13} this region actually could not be confirmed to be associated with rheumatoid arthritis. 

Four SNPs were then only detected by MOSGWA which are highlighted in the table above in yellow. These detections appear to be rather interesting. rs2104286 lies in the intron of the IL-2RA gene, and according to \cite{CV13} this SNP is definitely associated with rheumatoid arthritis. It was reported in \cite{WTCCC} after pooling data from RA and T1D, but it was not detected when using the standard analysis for data only from the RA population.
 rs1946518 from chromosome 11 is not mentioned in the meta-analysis, but one finds in the literature that the GUCY1A2 gene is down regulated in case of rheumatoid arthritis \citep{DI10}. This indicates that there might be a functional connection between this region and RA.

Finally there remain two SNPs from the GLI3 gene on chromosome 7. Again nothing is said about those in the meta-analysis of \cite{CV13}.  However, the GLI3 gene is known to be a member of the Hedgehog signaling pathway which is important for the proper development of embryos, and is also known to play an important role in adults \citep{VW00}. rs12536071 has a relatively small marginal p-value (5.2E-06), whereas the neighboring SNP rs12531052 has a rather large marginal p-values (0.167944 ). It is pretty unusual for MOSGWA to select two SNPs which are located so close to each other, which might indicate that actually some epistatic effect could be involved here.

\nopagebreak[4]

\subsection{Type I diabetes (T1D)}

\begin{table}[p]
\footnotesize{
\begin{tabular}{lrrlllllll}
dbSNP&Chr&Pos&Gene&M&HL&G3&G2&W\\ \hline
rs6679677 &1 &114303808 &RSBN1 &\textbullet &\textbullet &\textbullet &\textbullet & x\\
rs1025039 &5 &35901857 &CAPSL &\cy\textbullet & & & & \\
rs9261376  &\cb 6 &\cb 29.8-30.2MB&unknown &\textbullet & & & &\cb x  \\
rs16894900 &\cb 6 &\cb 29395499 &OR11A1 & &\textbullet & & &\cb x  \\
rs6906897  &\cb 6 &\cb 29415636 &OR12D3 &\textbullet & & & &\cb x  \\
rs17508548 &\cb 6 &\cb 29563067 &GABBR1 & &\textbullet & & &\cb x  \\
rs9258205 &\cb 6 &\cb 29703823 &LOC285830 &\textbullet & & & &\cb x  \\
rs1615251 &\cb 6 &\cb 29745761 &HCG4 & &\textbullet & & &\cb x  \\
rs3130531 &\cb 6 &\cb 31206616 &HLA-C &\textbullet &\textbullet & & &\cb x  \\
rs4959068 &\cb 6 &\cb 31343844 &MICA & &\textbullet & & &\cb x  \\
rs2248462 &\cb 6 &\cb 31446796 &MICB & & & &\textbullet &\cb x  \\
rs3131631 &\cb 6 &\cb 31484683 &MICB & &\textbullet &\textbullet &\textbullet &\cb x  \\
rs9348876 &\cb 6 &\cb 31575276 &NCR3 & &\textbullet & & &\cb x  \\
rs376510 &\cb 6 &\cb 31688200 &LY6G6C &\textbullet & & & &\cb x  \\
rs707915 &\cb 6 &\cb 31710968 &MSH5 & &\textbullet & & &\cb x  \\
rs2763979 &\cb 6 &\cb 31794592 &HSPA1B & &\textbullet &\textbullet &\textbullet &\cb x  \\
rs492899  &\cb 6 &\cb 31933518 &SKIV2L & &\textbullet & &\textbullet &\cb x  \\
rs408359  &\cb 6 &\cb 32141883 &AGPAT1 &\textbullet & & & &\cb x  \\
rs3129900 &\cb 6 &\cb 32305979 &C6orf10 & & & &\textbullet &\cb x  \\
rs3129933 &\cb 6 &\cb 32336161 &C6orf10 &\textbullet &\textbullet &\textbullet &\textbullet &\cb x  \\
rs3129934 &\cb 6 &\cb 32336187 &C6orf10 &\textbullet & &\textbullet &\textbullet &\cb x  \\
rs9391858 &\cb 6 &\cb 32341398 &C6orf10 &\textbullet & & & &\cb x  \\
rs2894254 &\cb 6 &\cb 32345689 &C6orf10 &\textbullet & & & &\cb x  \\
rs3135377 &\cb 6 &\cb 32385399 &HLA-DRA & & &\textbullet &\textbullet &\cb x  \\
rs3135393 &\cb 6 &\cb 32408842 &HLA-DRA & & & &\textbullet & \cb x \\
rs9268877 &\cb 6 &\cb 32431147 &HLA-DRB5 &\textbullet &\textbullet & &\textbullet &\cb x  \\ 
rs4530903 &\cb 6 &\cb 32581889 &HLA-DQA1 &\textbullet &\textbullet & &\textbullet &\cb x  \\
rs9272346 &\cb 6 &\cb 32604372 &HLA-DQA1 & & & &\textbullet &\cb x  \\
rs9272723 &\cb 6 &\cb 32609427 &HLA-DQA1 & &\textbullet &\textbullet &\textbullet &\cb x  \\
s9273363  &\cb 6 &\cb 32626272  &HLA-DQB1 &\textbullet &\textbullet &\textbullet &\textbullet &\cb x  \\
rs35120848&\cb 6 &\cb 32670495 &HLA-DQB1 & &\textbullet &\textbullet &\textbullet &\cb x  \\
rs2227127 &\cb 6 &\cb 32711782 &HLA-DQA2 & &\textbullet &\textbullet &\textbullet &\cb x  \\
rs2071474 &\cb 6 &\cb 32782582 &HLA-DOB & & & &\textbullet &\cb x  \\
rs241427 &\cb 6 &\cb 32804414 &TAP2 & & &\textbullet &\textbullet &\cb x  \\
rs3101942 &\cb 6 &\cb 32870057 &LOC100294145 & &\textbullet &\textbullet &\textbullet &\cb x  \\
rs3116985 &\cb 6 &\cb 33100590 &HLA-DPB2 &\textbullet & & & &\cb x  \\
rs7382464 &\cb 6 &\cb 33150268 &COL11A2 &\textbullet & & & &\cb x  \\
rs6928921 &6 &85062826 &KIAA1009 &\cy\textbullet & & & & \\
rs2666236 &10 &33418872 &NRP1 &\cy\textbullet & & & & \\
rs11171739 &12 &56470625 &ERBB3 &\textbullet &\textbullet & & &  x \\
rs17696736 &12 &112486818 &NAA25 &\textbullet &\textbullet & & & x \\
rs7157296 &14 &50566882 &C14orf138 &\cy\textbullet & & & & \\
rs12924729 &16 &11187783 &CLEC16A &\textbullet & & & & x \\
rs2542151 &18 &12779947 &PTPN2 &\cy\textbullet & & & & \\
rs41384747  &18 &16913046 &unknown &\cy\textbullet & & & & \\
\end{tabular}\\
}
\end{table}

Similarly to rheumatoid arthritis also for type I diabetes the HLA region plays an important role, where according to \cite{BQ11} approximately half of the genetic risk for T1D is found in HLA region.  GWASelect is handling the situation here slightly better than in case of RA, but still it detects apart from HLA SNPs only rs6679677 on chromosome 1, which is also reported by all other methods. In the original work \citep{WTCCC} three more SNPs are found to have strong association with T1D. All of those were found by MOSGWA, whereas HLASSO missed rs12924729 on chromosome 16. MOSGWA reported 6 additional SNPs which are again highlighted in yellow.

rs41384747 on chromosome 18 is close to rs2542151 which was already listed among SNPs with moderate association in  \cite{WTCCC}. The region was later confirmed in replication studies (see \cite{BC09}), and also includes SNP rs478582 listed in \cite{BQ11}. Similarly the region on chromosome 5 in which rs1025039 lies has been known as a susceptibility locus for T1D. In the same region lies rs6897932 which is associated with T1D according to \cite{BC09}. Perhaps the most comprehensive source today for human Type 1 Diabetes loci is the database \href{http://t1dbase.org}{http://t1dbase.org}. There one can find that the region on chromosome 18 is listed as being associated with T1D, but not the region on chromosome 5.

The closest region to rs6928921 on chromosome 6 which is documented in t1dbase is 6q15 with the BACH2 gene, which is 5 MB upstream. Nothing is known about any influence of KIAA1009 on T1D.  There is more indication that rs2666236 on chromosome 10 might be associated with T1D, as it is known that the corresponding gene NRP1 is associated with T1D \citep{HK10}. However, again the region of rs2666236 is not listed in t1dbase. Nothing is known about the other SNPs rs7157296 on chromosome 14 and rs41384747 on chromosome 18, and they might thus be false positives. Note that MOSGWA is tuned to have an FDR of approximately 10\%, and thus 2 or 3  false positive SNPs are to be expected in this model.

\subsection{Type II diabetes (T2D)}

\begin{tabular}{lrrlllllll}
dbSNP&Chr&Pos&Gene&M&HL&G3&G2&W\\ \hline
rs9465871 &6 &20717255 &CDKAL1 & & & &\textbullet & x \\
rs7917983 &\cb 10 &\cb 114732882 &TCF7L2 & & & &\textbullet & \cb x \\
rs7901275 &\cb 10 &\cb 114732906 &TCF7L2 & & & &\textbullet & \cb x \\
rs4074720 &\cb 10 &\cb 114748497 &TCF7L2 & &\textbullet & & & \cb x \\
rs7901695 &\cb 10 &\cb 114754088 &TCF7L2 & &\textbullet &\textbullet &\textbullet &\cb x  \\
rs4506565 &\cb 10 &\cb 114756041 &TCF7L2 &\textbullet & &\textbullet &\textbullet &\cb x  \\
rs7077039 &\cb 10 &\cb 114789077 &TCF7L2 & & &\textbullet &\textbullet & \cb x \\
rs7961581 &12 &71663102 &TSPAN8 & & & &\textbullet & \\
rs7193144 &\cb 16 &\cb 53810686 &FTO & &\textbullet & & & \cb x \\
rs8050136 &\cb 16 &\cb 53816275 &FTO &\textbullet & &\textbullet &\textbullet & \cb x \\
\end{tabular}\\

With type 2 diabetes a smaller number of genetic regions seems to be  associated than with the previous three diseases. Thus the model selection approach appears to have less benefits compared with the standard analysis. MOSGWA does not report rs9465871 on chromosome 6, but looking at the original analysis from \cite{WTCCC} this SNP has a trend p-value of 1.02 E-6, which is not significant after any standard correction for multiple testing. Apart from that all algorithms report SNPs within the two main regions on chromosome 10 and 16, where MOSGWA chooses again only one representative for gene TCF7L2 on chromosome 10.

\section{Discussion} \label{Sec:Discussion}

We have introduced MOSGWA, a new algorithm for GWAS analysis using the FDR controlling model selection criterion mBIC2. We compared its performance with two existing variable selection methods, GWASelect and HLASSO. The first observation was that both MOSGWA and HLASSO are controlling the Type I error rate under the global null hypothesis, whereas GWASelect does not manage to do that when using the recommended parameter setting $\xi \in [0.1, 0.2]$. \cite{He11} only presented simulation results for scenarios including 10 causal SNPs, for which they reported relatively low FDR. In accordance we observed that GWASelect tends to have lower type I error rate when the true model underlying simulations includes more causal SNPs. 
However, a method which selects very large models even when we know that there is no genetic cause for the disease status appears to be rather problematic. When increasing the selection threshold to $\xi = 0.3$ then GWASelect more or less controls the type I error under the global null, but then it is no longer competitive in terms of power to detect causal SNPs. One advantage of MOSGWA is that there is actually no parameter tuning necessary at all, because the selection criterion is fixed.

Apart from GWASelect and HLASSO we have also compared the performance of the original LASSO as implemented in glmnet \cite{FHT10}, although we decided not to present the corresponding results in this manuscript. LASSO previously has been shown to perform well in GWAS in terms of prediction \cite{Koo}, but here we are only interested in selecting the correct SNPs. For this purpose one has the problem to decide upon the best tuning parameter $\lambda$ of LASSO. It is well known that cross validation yields too large models, which we also observed in our simulations. As an alternative we tried to search along the LASSO regularization path and find the model which minimizes mBIC2, but that gave too small models. We finally considered a strategy where we searched along the regularization path for that model which minimizes the misclassification rate. This is obviously not  feasible in practice when the truth is unknown, but this strategy shows the best possible performance of LASSO that one could achieve at least theoretically. It still turned out that in our simulations even the best possible model along the regularization path could not compete with the models obtained with MOSGWA or  HLASSO.  We observed that LASSO tends to select too many correlated SNPs with larger effect sizes, and then has difficulties to include causal SNPs with smaller effect size, but rather includes a number of false positives.
An explanation of this behavior was given already by \cite{Hog8} when motivating the NEG prior.

Our simulation study on complex traits showed that at least for our three scenarios MOSGWA is slightly more powerful than HLASSO, when the parameter $\alpha$ of HLASSO is chosen such that both procedures have similar type 1 error rate. This reflects the theoretical optimality property of mBIC to minimize asymptotically the misclassification error under a wide range of sparsity levels \citep{FBMC}. However, this theoretical property holds for the model which actually minimizes the criterion mBIC2. Our heuristic search strategy is attempting to get close to the global minimum, but we know that in most cases it will fail to find the best solution.  More involved search strategies will further improve our method, and we are currently exploring the us of memetic algorithms, which have been successfully applied already in the context of QTL mapping \cite{FLAB}.

The software architecture of MOSGWA is designed in such a way that it can be quite easily extended in the future to incorporate more advanced features. For example it might be interesting to have a model selection procedure which accounts for population structure. Currently this can be done by adding principle components as covariates to the regression models, but an even better solution would be to add random effects to model population structure. For that reason we are currently  working on extending MOSGWA  towards mixed models.

%
%

\section*{Acknowledgment}

{\bf Funding} This research has been funded by the Vienna Science and Technology Fund
(WWTF) through project MA09-007a.

%
%

\end{document}